\documentclass[aps,prd,reprint,showpacs,floatfix,longbibliography,nofootinbib,superscriptaddress]{revtex4-1}

\usepackage{tikz,graphicx,booktabs,multirow,array,microtype,mathtools,float}

\usepackage{centernot}
\newcommand{\fsl}[1]{{\centernot{#1}}} % Feynman slash

\graphicspath{{figures/}}
\usepackage{enumerate}   
\extrafloats{200}
\usepackage[colorlinks=true,backref=false, linktocpage=true,
citecolor=blue,
urlcolor=blue,linkcolor=blue,
pdfpagemode=UseOutlines]{hyperref}
\newcolumntype{C}{>{$}c<{$}}
\usepackage{dcolumn}

\usepackage{anyfontsize}
\usepackage[shortlabels]{enumitem}

\usepackage{amsfonts}
\AtBeginDocument{
\heavyrulewidth=.08em
\lightrulewidth=.05em
\cmidrulewidth=.03em
\belowrulesep=.65ex
\belowbottomsep=0pt
\aboverulesep=.4ex
\abovetopsep=0pt
\cmidrulesep=\doublerulesep
\cmidrulekern=.5em
\defaultaddspace=.5em
}
\newcommand{\langl}{\begin{picture}(4.5,7)
\put(1.1,2.5){\rotatebox{60}{\line(1,0){5.5}}}
\put(1.1,2.5){\rotatebox{300}{\line(1,0){5.5}}}
\end{picture}}
\newcommand{\rangl}{\begin{picture}(4.5,7)
\put(.9,2.5){\rotatebox{120}{\line(1,0){5.5}}}
\put(.9,2.5){\rotatebox{240}{\line(1,0){5.5}}}
\end{picture}}

\usepackage{amsmath} % wonderful math package
\usepackage{dsfont}  % double strike font
\usepackage{bm}      % bold math

\def\beq{\begin{equation}}
\def\eeq{\end{equation}}
\def\beqs#1\eeqs{\beq\begin{split} #1 \end{split}\eeq}

\long\def\comment#1{}

\begin{document}

\title{
%Towards a lattice QCD determination of hadron polarizabilities from four-point functions\\
Towards charged hadron polarizabilities from four-point functions in lattice QCD
}% Force line breaks with \\

\author{Walter Wilcox}
\email{Walter\_Wilcox@baylor.edu}
\affiliation{Department of Physics, Baylor University, Waco, Texas 76798, USA}
\author{Frank X. Lee}
\email{fxlee@gwu.edu}
\affiliation{Physics Department, The George Washington University, Washington, D.C. 20052, USA}
\date{\today} 
%\date{xx xx, 2021} 

\begin{abstract}
We show how to compute electromagnetic polarizabilities of charged hadrons using four-point functions in lattice QCD. 
The low-energy behavior of Compton scattering amplitude is matched to matrix elements of current-current correlation functions on the lattice. Working in momentum space, formulas for electric polarizability ($\alpha_E$) and magnetic polarizability ($\beta_M$) are derived for both charged pion and proton. 
Lattice four-point correlation functions are constructed from quark and gluon fields to be used in Monte Carlo simulations. The content of the functions is assessed  in detail and specific prescriptions are given to isolate the polarizabilities. The connected quark-line diagrams can be done today as a small lattice project. The disconnected diagrams are more challenging but are within reach of dedicated resources for medium to large lattice projects. We also draw attention to the potential of four-point functions as a multipurpose tool for hadron structure.
\end{abstract}

\maketitle

\section{\label{sec:intro}Introduction}

Electromagnetic polarizabilities are important properties that shed light on the internal structure of hadrons. The quarks respond to probing electromagnetic fields, revealing the charge and current distributions inside the hadron. 
There is an active community in nuclear physics partaking in this endeavor.
Experimentally, polarizabilities are primarily studied by low-energy Compton scattering. On the theoretical side, a variety of methods have been employed to describe the physics involved, from phenomenological models~\cite{Lvov:1993fp,Lvov_2001}, to chiral perturbation theory (ChPT)~\cite{Moinester:2019sew,Lensky:2009uv,Hagelstein:2020vog} or chiral effective field theory (EFT)~\cite{McGovern:2012ew,Griesshammer:2012we}, 
to lattice QCD. Reviews of the experimental status can be also found in Refs.~\cite{Moinester:2019sew,Griesshammer:2012we}.

Understanding electromagnetic polarizabilities has been a long-term goal of lattice QCD. The challenge lies in the need to apply both QCD and QED principles. The standard tool to compute polarizabilities is the background field method which has been widely used~\cite{Fiebig:1988en,Lujan:2016ffj, Lujan:2014kia, Freeman:2014kka, Freeman:2013eta, Tiburzi:2008ma, Detmold:2009fr, Alexandru:2008sj, Lee:2005dq, Lee:2005ds,Engelhardt:2007ub,Bignell:2020xkf,Deshmukh:2017ciw,Bali:2017ian,Bruckmann:2017pft,Parreno:2016fwu,Luschevskaya:2015cko,Chang:2015qxa,Detmold:2010ts}.  Methods to study higher-order polarizabilities have also been proposed~\cite{Davoudi:2015cba,Engelhardt:2011qq,Lee:2011gz,Detmold:2006vu} in this approach. 
Although such calculations are relatively straightforward, requiring only two-point functions, there are a number of unique challenges.
First, since weak fields are needed, the energy shift involved is very small relative to the mass of the hadron (on the order of one part in a million depending on field strength). This challenge has been successfully overcome by relying on statistical correlations with or without the field. 
Second, there is the issue of discontinuities across the boundaries when applying a uniform field on a periodic lattice. This has been largely resolved by using quantized values for the fields.
Third and more importantly, a charged hadron accelerates in an electric field and exhibits Landau levels in a magnetic field. Such motions are unrelated to polarizability and must be isolated from the deformation due to quark and gluon dynamics inside the hadron. For this reason, most calculations have focused on neutral hadrons. Since the standard plateau technique of extracting energy from the large-time behavior of the two-point correlator fails for charged hadrons, special techniques are needed to filter out the collective motion of the system in order to extract polarizabilities~\cite{niyazi2021charged,Bignell_2020,He:2021eha,Detmold:2009fr}.

In this work, we examine the use of four-point functions to extract polarizabilities. 
As we shall see, the method is ideally suited to charged hadrons; there is no background field to speak of.
Furthermore, the method directly mimics the Compton scattering process on the lattice. 
Although four-point correlation functions have been applied to various aspects of hadron structure~\cite{Liang:2019frk,Liang_2020a,Fu_2012,Alexandrou_2004,Bali_2018,bali2021double},  
not too much attention has been paid to its potential application for polarizabilities.
The only work we are aware of are two attempts 25 years ago, one based in position space~\cite{BURKARDT1995441} and one in momentum space~\cite{Wilcox:1996vx}. 
Here, we want to take a fresh look at the problem. 

In Sec.~\ref{sec:pion} we outline the methodology using electric polarizability of charged pion as an example, then extend it to magnetic polarizability. In Sec.~\ref{sec:proton} 
we derive new formulas for both electric and magnetic polarizabilities of the proton. 
In Sec.~\ref{sec:lat} we detail how to measure the polarizabilities by  
constructing QCD four-point correlators on the lattice, 
and describe methods to isolate them. Concluding remarks are in Sec.~\ref{sec:con}.

%%%%%%%%%%%%%%%%%%%%%%%%%%%%%
\vspace*{-3mm}
\section{Charged pion}
\label{sec:pion} 
The pion is the simplest hadronic system to demonstrate the methodology.
First we briefly review the essential steps to connect electric polarizability 
to lattice matrix elements of four-point functions.
Then we extend the method to derive the formula for magnetic polarizability. 

%-----------------------
\subsection{Electric polarizability}
\label{sec:alpha} 

\begin{figure}[h]
\includegraphics[scale=0.55]{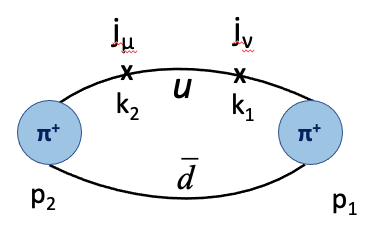}
\caption{Pictorial representation of the four-point function in Eq.\eqref{eq:4pt} for $\pi^+$ (for protons imagine two $u$ and one $d$ quark lines). 
Time flows from right to left and the four-momentum conservation is $p_2 +k_2 = k_1+p_1$.}
\label{fig:diagram-4pt1}
\end{figure}

For this part, we follow closely the notations and conventions of Ref.~\cite{Wilcox:1996vx}.
The central object is the time-ordered Compton scattering tensor 
defined by the four-point correlation function,\footnote{We use round brackets $(\cdots |\cdots)$ to denote continuum matrix elements, and angle brackets $\langl\cdots |\cdots\rangl$ to denote lattice matrix elements.}
\beq
T_{\mu\nu}=i\int d^4x e^{ik_2\cdot x}( \pi(p_2)|T j_\mu(x) j_\nu(0) | \pi(p_1))
\label{eq:4pt}
\eeq
where the electromagnetic current density is
\beq
j_\mu=q_u \bar{u}\gamma_\mu u + q_d \bar{d}\gamma_\mu d,
\label{eq:current}
\eeq
built from up and down quark fields ($q_u=2/3$, $q_d=-1/3$).
The function is represented in Fig.~\ref{fig:diagram-4pt1}.
We work with a special kinematical setup called a zero-momentum Breit frame given by,
\beqs
p_1 &=(m,\vec{0}),\\
k_1 &=(0,\vec{k}),\;
k_2 =(0, \vec{k}), \; \vec{k}=(0,0,k_z),\;  k_z \ll m,\\
p_2 &=-k_2+k_1+p_1=(m,\vec{0}).
\label{eq:Breit1}
\eeqs
Essentially it can be regarded as forward double virtual Compton scattering. This is different from the real Compton scattering in experiments. They access the same low energy constants including the polarizabilities.

On the phenomenological level, the process can be described by an effective relativistic theory to expose its physical content. 
The tensor can be parametrized to second order in photon momentum by the general  form,
\beqs
&\sqrt{2E_12E_2} \,T_{\mu\nu} = \\
& -{T_\mu(p_1+k_1,p_1) T_\nu(p_2,p_2+k_2) \over (p_1+k_1)^2-m^2 } \\
& -{T_\mu(p_2,p_2-k_1) T_\nu(p_1-k_2,p_1) \over (p_1-k_2)^2-m^2 } +2g_{\mu\nu} \\
&+ A(k_1^2g_{\mu\nu} - k_{1\mu}k_{1\nu} + k_2^2 g_{\mu\nu} - k_{2\mu}k_{2\nu}) \\
&+ B(k_1\cdot k_2 g_{\mu\nu} -  k_{2\mu}k_{1\nu}) \\
&+ C(k_1\cdot k_2 Q_\mu Q_\nu + Q\cdot k_1 Q\cdot k_2  g_{\mu\nu} \\
&\quad\quad - Q\cdot k_2 Q_\mu k_{1\nu} - Q\cdot k_1 Q_\nu k_{2\mu}),
\label{eq:Tmn}
\eeqs 
where $Q=p_1+p_2$ and $A$, $B$, $C$ are constants to be characterized.~\footnote{ We use a Minkowski metric $g_{\mu\nu}=\text{diag}(1,-1,-1,-1)$. For example, the relativistic energy-momentum relation for 4-momentum $p^\mu=(E,\vec{p})$  is $p^2=p^\mu g_{\mu\nu}p^\nu=E^2-\vec{p}^{\,2}=m^2$.}
We use a noncovariant normalization 
\beq
\sum_n\int {d^3p\over (2\pi)^3} |n(p))( n(p)|=1,
\eeq
which is why the square root factor in front of $T_{\mu\nu} $.
The pion electromagnetic vertex with momentum transfer $q=p'-p$ is written as 
\beq
T_\mu(p',p)=(p'_\mu+p_\mu) F_\pi(q^2) +q_\mu {p'^2-p^2\over q^2}(1-F_\pi(q^2)).
\eeq
It satisfies $q_\mu T_\mu(p',p)=p'^2-p^2$ for off-shell pions, which is needed for the Ward-Takahashi identity.
The pion form factor to 4th order in momentum is given by 
\beq
 F_\pi(q^2) =1 + {\langl r^2\rangl \over 6} q^2 +  {\langl r^4\rangl \over 120} q^4, 
 \label{eq:pionff}
\eeq
where $\langl r^2\rangl$ is the pion squared charge radius and $q^2<0$ is space-like 4-momentum transfer squared. The form in Eq.\eqref{eq:Tmn} can be entirely motivated by general principles of Lorentz invariance, gauge invariance, current conservation, time-reversal symmetry, and crossing
symmetry~\cite{Moinester:2019sew}.
In fact, current conservation ($k^\mu_1T_{\mu\nu}= k^\nu_2T_{\mu\nu}=0$) immediately leads to $A$ being related to charge radius by $A= {\langl r^2\rangl / 3}$.
The first three terms in Eq.\eqref{eq:Tmn}  are the Born contributions to scattering from the pion and the remaining three are contact terms.
To make contact with polarizabilities, we need to take the nonrelativistic limit of  the Compton scattering amplitude 
$\alpha \epsilon^\mu_1T_{\mu\nu} \epsilon^{\nu*}_2$
where $\epsilon_1$ and $\epsilon_2$ are the initial and final photon polarization 4-vectors.
We use a unit system in which the fine structure constant $\alpha=e^2 \simeq 1/137$ where the unit charge $e$ has been pulled from Eq.\eqref{eq:current}. 
%That the square of the charge appears in the four-point function immediately indicates that $\pi^+$ and $\pi^-$ have the same polarizabilities.
For the purpose of a nonrelativistic limit, we have the freedom to choose the following kinematics:
The initial pion is at rest in the lab frame;  photons have spatial polarization unit vectors.
Then, the relativistic Compton scattering amplitude 
can be reduced to the standard form, to quadratic order in photon energy and momentum, 
\beqs
&\alpha \epsilon^{\mu}_1 T_{\mu\nu} \epsilon^{\nu*}_2 = \\ 
&\hat{\epsilon}_1\cdot \hat{\epsilon}_2^* 
 \left[ -{\alpha\over m} \left(1+{\langl r^2\rangl \over 6} (k_1^2 + k_2^2)\right) 
+\alpha_E \omega_1\omega_2 \right] \\
&+ \beta_M (\hat{\epsilon}_1\times \vec{k}_1)\cdot (\hat{\epsilon}^*_2\times \vec{k}_2), 
\label{eq:low}
\eeqs
where the $\alpha/m$ term is the Thomson limit corresponding to the $g_{\mu\nu}$ term, the $\langl r^2\rangl$ corresponds to the $A$ term, and the $\alpha_E$ and $\beta_M$ terms come from $B$ and $C$, which 
 are related to the polarizabilities by
\beqs
\alpha_E &\equiv -\alpha\left( {B\over 2m} + 2mC \right),\\
\beta_M &\equiv  \alpha {B\over 2m}.
\label{eq:BC}
\eeqs
So $B$ is a proxy for magnetic polarizability, and $C$ is the sum of electric and magnetic polarizabilities.
All three parameters in Eq.\eqref{eq:Tmn} are now identified with physical meanings.

For electric polarizability, we work with the $\mu=\nu=0$ component of Eq.\eqref{eq:Tmn}.
Under the special kinematics in Eq.\eqref{eq:Breit1}, it can be written to order $\vec{k}^{\,2}$ in the form,
\beqs
&T_{00}(\vec{k}) = {4m_\pi \over \vec{k}^{\,2}} + \left( {1\over m_\pi} - {4\over 3}m_\pi\langl r^2\rangl\right)\\
&+ \left[ - {\langl r^2\rangl \over 3 m_\pi}+ {1\over 9}m_\pi \langl r^2\rangl^2 + {1\over 15}m_\pi \langl r^4\rangl + {\alpha^\pi_E\over \alpha}  \right]  \vec{k}^{\,2}\\
& \equiv T^{Born}_{00}(\vec{k}) + {\alpha^\pi_E\over \alpha} \vec{k}^{\,2},
\label{eq:T00_pion}
\eeqs
where we separate the Born contribution from the contact term.

The next step is to relate the polarizabilities to lattice matrix elements.
To this end, we need to convert from continuum to a lattice of isotropic spacing $a$ with $N_s=N_x\times N_y\times N_z$  number of spatial sites  by the following correspondence,
\beqs
|n(p)) &\to V^{1/2} |n(p)\rangl, \\
j_\mu(x) &\to {Z_V \over a^3} j^{L}_\mu(x), \\
\int d^4x &\to a^4 \int_{-\infty}^\infty dt \sum_{\vec{x}},
 \eeqs
where $V=N_s a^3$ and the superscript $L$ denotes  they are lattice versions of the continuum entities.  We are still in Minkowski spacetime for the purpose of matching with the continuum. We also keep the time continuous but dimensionless. The renormalization factor $Z_V$ for the lattice current $j^{L}_\mu=(\rho^L,-\vec{j}^L)$ can be taken to be unity if conserved currents are used on the lattice.
Equation~\eqref{eq:4pt} becomes,
\beq
T_{\mu\nu}=i\ N_s a \int dt \sum_{\vec{x}} e^{ik_2\cdot x} \langl \pi(p_2) |T j^L_\mu(x) j^L_{\nu}(0) | \pi(p_1)\rangl.
\label{eq:4ptlat}
\eeq
On the lattice, there is a contribution to this function when $p_1=p_2$, called a vacuum expectation value (or VEV), that must be subtracted out. The reason is we are interested in differences relative to the vacuum, not the vacuum itself.
Formally, this is enforced by requiring normal ordering instead of time ordering in Eq.\eqref{eq:4ptlat},
\beq
:j^L_\mu(x) j^L_\nu(0) : = T j^L_\mu(x) j^L_\nu(0)  - \langl 0|T j^L_\mu(x) j^L_\nu(0) | 0\rangl.
\eeq

For electric polarizability, the relevant component is $T_{00}$ which amounts to the overlap of charge densities.  
By inserting a complete set of intermediate states,
making use of translation invariance of the lattice current,
and integrating over time, we arrive at the subtracted correlator,~\footnote{In this work we use $\vec{k}$ to denote continuum momentum and $\vec{q}$ lattice momentum with the same physical unit. When we match the two forms we set $\vec{k}=\vec{q}$ and express the result in terms of $\vec{q}$.}
\beqs
T_{00} &= 
2N_s^2 \sum_n { |\langl\pi(\vec{0})| \rho^L(0) | n(\vec{q})\rangl|^2 \over E_n-m_\pi }\\
&\quad - 2N_s^2 \sum_n { |\langl 0 | \rho^L(0) | n(\vec{q})\rangl|^2 \over E_n }\\
& \equiv T^{elas}_{00} + T^{inel}_{00},
\label{eq:T00_pion_lat}
\eeqs
where the elastic part ($n=\pi$ term in the first sum) is separated from the inelastic part as, 
\beq
T^{elas}_{00} \equiv 
2N_s^2 { |\langl\pi(\vec{0})| \rho^L(0) | \pi(\vec{q}\rangl|^2 \over E_\pi-m_\pi }.
\eeq
The matrix element 
\beq
\langl\pi(\vec{0})| \rho^L(0) | \pi(\vec{q})\rangl={1\over N_s} {E_\pi+m_\pi\over \sqrt{2E_\pi 2m_\pi}} F_\pi(q^2),
\eeq
is related to the pion form factor $F_\pi$  given in Eq.\eqref{eq:pionff}.
Setting $q^2=2m_\pi(m_\pi-E_\pi)$ and evaluating $T_{00}^{elas}$ to order $\vec{q}^{\,2}$, it takes the form,
\beqs
&T_{00}^{elas}(\vec{q})  = {4m_\pi \over \vec{q}^{\,2}} + \left( {1\over m_\pi} - {4\over 3}m_\pi\langl r^2\rangl\right)\\
&+ \left[ {1\over 9}m_\pi \langl r^2\rangl^2 + {1\over 15}m_\pi \langl r^4\rangl \right]  \vec{q}^{\,2}.
\eeqs

The final step is to match the effective lattice $T_{00}$ given in Eq.\eqref{eq:T00_pion_lat} with the same component in the continuum given in Eq.\eqref{eq:T00_pion} under the same kinematics  and to the same order,
\beq
T_{00} =  T^{Born}_{00}(\vec{q}) + {\alpha_E \over \alpha}  \vec{q}^{\,2}.
\eeq
Subtracting off $T_{00}^{elas}$ from this equation, we get
\beq
T_{00}(\vec{q})-T_{00}^{elas}(\vec{q})  =  -\frac{\langl r^2\rangl}{3m_{\pi}}\vec{q}^{\,2} + {\alpha_E \over \alpha}  \vec{q}^{\,2},
\eeq
or a formula for charged pion electric polarizability on the lattice,
\beq
\alpha^\pi_E=  \alpha \left[ \frac{\langl r^2\rangl}{3m_{\pi}} +\frac{T_{00}(\vec{q}_1) - T_{00}^{elas}}{\vec{q}_1^{\,2}}\right],
\label{eq:alpha_pion}
\eeq
where $\vec{q}_1$ emphasizes that the formula is valid for the smallest nonzero spatial momentum on the lattice.
The physical unit checks out to be $a^3$ (fm$^3$) since $\vec{q}_1^2 $ has the unit of $a^{-2}$ and $T_{00}$ has the unit of $a$ (or $1/m_\pi$).
We discuss how to measure it on the lattice in a later section.
For now, it suffices to say that charged pion's 
$\alpha_E$ can be positive or negative in a relativistic quantum field theory, since it has the difference of two positive-definite quantities in Eq.\eqref{eq:T00_pion_lat}.

%-----------------------
\subsection{Magnetic polarizability}
\label{sec:beta} 

Magnetic polarizability proceeds in a similar fashion, except we consider the spatial component $T_{11}$ ($T_{22}$ gives the same result). 
Under the same kinematics given  in Eq.\eqref{eq:Breit1}, 
this component from the general form in Eq.\eqref{eq:Tmn} reads
\beq
2m_\pi T_{11}= -2 + {2\langl r^2\rangl \over 3} \vec{k}^{\,2} + B\, \vec{k}^{\,2},
\eeq
or 
\beq
T_{11}= -{1\over m_\pi} + \vec{k}^{\,2} \left( {\langl r^2\rangl \over 3m_\pi} + {\beta_M \over \alpha} \right).
\label{eq:4pt11}
\eeq
On the other hand, from the lattice four-point function in Eq.\eqref{eq:4ptlat}, we have,
\beq
T_{11}=i\ N_s a \int_{-\infty}^\infty dt \sum_{\vec{x}} e^{ik_2\cdot x}\langl \pi(p_2)|T j^L_1(x) j^L_1(0) |\pi(p_1)\rangl.
\eeq
Here we examine its context in more detail. 
Similar steps were used in the electric case~\cite{Wilcox:1996vx}.
The first thing to do is to apply the same zero-momentum Breit frame kinematics.
Then, we insert a complete set of intermediate states in between the currents,
\beq
\sum_{n,\vec{p}_n,s}  | n(\vec{p}_n,s)\rangl\langl n(\vec{p}_n,s) |=1,
\eeq
where $s$ is the spin label for the intermediate state.
The translation invariance of $j^L_1(x)$ gives,
\beqs
j^L_1(x) &= e^{ip_\pi\cdot x} j^L_1(0) e^{-ip_n\cdot x} \\
&= e^{im_\pi t a} j^L_1(0) e^{-i(E_n a t - \vec{p}_n \cdot \vec{x} )},
\eeqs
where the second step makes explicit the result of acting on the zero-momentum state on the left and the on-shell intermediate state on the right.
The spatial sum can be collapsed with the delta function ($\vec{k}_2=\vec{q}$),
\beq
 \sum_{\vec{x}} e^{i(-\vec{q}+\vec{p}_n)\cdot \vec{x}} = \delta_{\vec{q},\vec{p}_n} N_s.
\eeq
The principal value of the time integral is,
\beq
\int_{-\infty}^\infty dt \;e^{ia(m_\pi -E_n ) t} ={ 2 \over i a (E_n-m_\pi)}.
\eeq
These steps, together with VEV subtraction, lead to
\beqs
T_{11}(\vec{q})&=
2N_s^2 \sum_{n,s} { |\langl\pi(\vec{0})| j^L_1(0) | n(s,\vec{q}\rangl|^2 \over E_n-m_\pi } \\
& - 2 N_s^2  \sum_{n,s} { |\langl 0| j^L_1(0) | n(s,\vec{q}\rangl|^2 \over E_n }.
\label{eq:T11}
\eeqs
Note that the elastic piece ($n=\pi$) in the sum vanishes under the special kinematics, 
\beq
\langl\pi(\vec{0})| j^L_1(0) | \pi(\vec{q},s)\rangl=0.
\eeq
The reason is that the matrix element is proportional to $(\vec{0}+\vec{q})_1$ in 1-direction but momentum $\vec{q}$ is in 3-direction.

For the inelastic contributions, the types of intermediate state contributing are vector or axial vector mesons~\cite{Wilcox:1996vx}. There is no need to analyze the matrix elements explicitly as done in Ref.~\cite{Wilcox:1996vx} for the electric case. We only need to know that the inelastic part can be characterized up to order $\vec{q}^{\,2}$ by the form,
\beq
T_{11}(\vec{q}) \equiv T_{11}(\vec{0}) + \vec{q}^{\,2}  K_{11},
\label{eq:inelastic_lat}
\eeq
with $T_{11}(\vec{0})$ and $K_{11}$ to be related to physical parameters and determined on the lattice.
Note that we deliberately use the full amplitude label $T_{11}$ instead of $T^{inel}_{11}$ since the elastic part is zero.

Matching the full amplitude on the lattice in Eq.\eqref{eq:4pt11} with the continuum version in Eq.\eqref{eq:inelastic_lat}, 
we obtain two relations,
\begin{align}
 -{1\over m_\pi}  &= T_{11}(0), \label{eq:sr} \\
  {\langl r^2\rangl \over 3m_\pi} + {\beta_M \over \alpha}  &= K_{11}.  \label{eq:K11}
  \end{align}
The first relation is a sum rule at zero momentum.
The second leads to a formula for charged pion magnetic polarizability, 
\beq
\beta^\pi_M= \alpha \left[- {\langl r^2\rangl \over 3m_\pi} + { T_{11}(\vec{q}_1) -  T_{11}(0)\over \vec{q}_1^{\,2}} \right],
\label{eq:beta_pion}
\eeq
where  $\vec{q}_1$ is the lowest momentum on the lattice.
Compared to charged pion electric polarizability $\alpha^\pi_E$ in Eq.\eqref{eq:alpha_pion}, we see that instead of subtracting the elastic contribution, we subtract the zero-momentum inelastic contribution in the magnetic polarizability. 
In other words, there is no zero-momentum contribution in $\alpha^\pi_E$, and no elastic contribution in $\beta^\pi_M$.

The zero-momentum sum rule in Eq.\eqref{eq:sr} warrants more discussion.
Isolating it from Eq.\eqref{eq:T11}, this term reads,
\beqs
T_{11}(\vec{0})&=
2N_s^2 \sum_{n,s} { |\langl\pi(\vec{0})| j^L_1(0) | n(s,\vec{0}\rangl|^2 \over m_n-m_\pi } \\
& - 2 N_s^2  \sum_{n,s} { |\langl 0| j^L_1(0) | n(s,\vec{0}\rangl|^2 \over m_n }.
\label{eq:T11zero}
\eeqs
It plays the role of an inelastic subtraction in the determination of $\beta^\pi_M$. The first term represents the inelastic contribution and is positive definite, whereas the second term represents the VEV contribution and is negative definite.
The fact that they sum to a negative value ($T_{11}(\vec{0})=-1/m_\pi$) signifies that the VEV associated contributions dominate at zero momentum.
In the valence approximation, defined as dropping all disconnected loops which include diagrams (d), (e), and (f) in Fig.\ref{fig:diagram-4pt3} below and the VEVs associated with them, we expect $T^{val}_{11}(0)>0$ since the only contribution left is from the first term in Eq.\eqref{eq:T11zero}. 
In a full calculation, we do expect $T_{11}(\vec{0})<0$,  and the sum rule can be used as a guide on the sign of $T_{11}(\vec{0})$. We do not expect however to use it to determine the mass. 

%%%%%%%%%%%%%%%%%%%%%%%%%%%%%
\section{Proton}
\label{sec:proton} 
As the simplest nucleus, a proton's polarizabilties are of fundamental importance to our understanding of structure of matter.
They are more precisely measured than a charged pion's polarizabilties in Compton scattering experiments on hydrogen targets.
Theoretically, chiral perturbation theory is well established on the proton. 
The four-point function lattice QCD approach considered here offers a much-needed addition to the effort.
The formalism parallels that for a charged pion, 
but is more complicated mainly due to a proton's spin-1/2 structure. 
We first consider electric polarizability in sufficient detail, then build it out to magnetic polarizability.

%-----------------------
\subsection{Electric polarizability}

We start with a general proton Compton tensor parametrized to second order in photon momentum,
\beqs
\sqrt{2E_12E_2} \,T_{\mu\nu} &= T_{\mu\nu}^{Born}
+ B(k_1\cdot k_2 g_{\mu\nu} -  k_{2\mu}k_{1\nu}) \\
&+ C(k_1\cdot k_2 Q_\mu Q_\nu + Q\cdot k_1 Q\cdot k_2  g_{\mu\nu} \\
&\quad\quad - Q\cdot k_2 Q_\mu k_{1\nu} - Q\cdot k_1 Q_\nu k_{2\mu}),
\label{eq:Tmn_p}
\eeqs 
where $Q=p_1+p_2$.
For the Born term we take from Ref.~\cite{gasser2020cottingham},
\beqs
T_{\mu\nu}^{Born} &= {B_{\mu\nu}(p_2,k_2,s_2 | p_1,k_1,s_1) \over m_p^2 - s} \\
&+ {B_{\nu\mu}(p_2,-k_1,s_2 | p_1,-k_2,s_1) \over m_p^2 - u},
\eeqs
where the function is (note a factor of 1/2 difference between our definition and Ref.~\cite{gasser2020cottingham}),
\beqs
& B_{\mu\nu}(p_2,k_2,s_2 | p_1,k_1,s_1) =  \\
&\bar{u}(p_2,s_2) \Gamma_{\mu}(-k_2) (\fsl{P}+m_p) \Gamma_{\nu}(k_1)  u(p_1,s_1).
\eeqs
Here $P=p_2+k_2=p_1+k_1$ is the standard 4-momentum conservation for Compton scattering. 
There is no $A$ term here because the proton Born terms obey current conservation, unlike the pion case in Eq.\eqref{eq:Tmn}.
A nonrelativistic reduction of Eq.\eqref{eq:Tmn_p} has the same form as Eq.\eqref{eq:low} for the pion, except for the noncontact $\langl r^2\rangl$ term, with the same $B$ and $C$ relations to polarizabilities as in Eq.\eqref{eq:BC}.

The Born amplitude has virtual (or off-shell) intermediate hadronic states in the s and u channels, whereas on the lattice we have real (or on-shell) intermediate states. This will produce a difference with the elastic contribution to be discussed later.
The vertex function is defined by
\beq
 \Gamma_{\mu}(q)\equiv \gamma_\mu F_1(q) + {i F_2(q)\over 2m_p} \sigma_{\mu\lambda} q^\lambda,
\eeq
where $q=p'-p$ is the 4-momentum transfer at the vertex and  summation over $\lambda$ is implied. 
Specializing to our kinematics in Eq.\eqref{eq:Breit1}, we have
\beqs
s&=(p_1+k_1)^2=m_p^2-\vec{k}^{\,2}, \\
u&=(p_1-k_2)^2=m_p^2-\vec{k}^{\,2}. 
\eeqs
We consider an unpolarized Born expression given by
\begin{align}
&2m_p T_{\mu\nu}^{Born} ={1\over 2} \sum_{s_1,s_2} {1\over \vec{k}^{\,2}}  \label{eq:TB} \\
&\left[
\bar{u}(\vec{0},s_2) 
\left(\gamma_\mu F_1 - {iF_2\over 2m_p}\sigma_{\mu\lambda} k_2^\lambda\right) \right. \nonumber \\
& \left. (\fsl{p_1}+\fsl{k_1}+m_p) 
\left(\gamma_\mu F_1 + {iF_2\over 2m_p}\sigma_{\mu\lambda} k_1^\lambda\right)
u(\vec{0},s_1)  \right. \nonumber \\
& \left. + \bar{u}(\vec{0},s_2) 
\left(\gamma_\nu F_1 + {iF_2\over 2m_p}\sigma_{\nu\lambda} k_1^\lambda\right)  \right. \nonumber \\
& \left. (\fsl{p_1}-\fsl{k_2}+m_p)  
\left(\gamma_\nu F_1 - {iF_2\over 2m_p}\sigma_{\nu\lambda} k_2^\lambda\right)
u(\vec{0},s_1)
\right]. \nonumber
\end{align}
 The zero-momentum spinors are given by,
\beq
u(\vec{0},1)= \sqrt{2m_p} \left(\begin{array}{c} 1 \\ 0 \\ 0 \\ 0 \end{array} \right), 
u(\vec{0},2)= \sqrt{2m_p} \left(\begin{array}{c} 0 \\ 1 \\ 0 \\ 0 \end{array} \right),
\eeq
using our normalization, and $\bar{u}\equiv u^\dagger \gamma^0$.
For gamma matrices, we use the standard Dirac basis,
\beqs
&\gamma^0= \left( \begin{array}{cc} I & 0 \\   0 & -I \end{array} \right), 
\gamma^i= \left(\begin{array}{cc} 0 & \sigma_i \\   -\sigma_i & 0 \end{array} \right),\\
&\gamma^\mu\gamma^\nu + \gamma^\nu\gamma^\mu =2g^{\mu\nu}=2g_{\mu\nu}, \\
&\sigma^{\mu\nu}={i\over 2} \left(\gamma^\mu\gamma^\nu - \gamma^\nu\gamma^\mu \right).
\eeqs

Next, we need to express this amplitude as a series in $\vec{k}^{\,2}$.
To this end, we choose to expand the Sachs form factors first.
Since the Born term has a $1/\vec{k}^{\,2}$ pole, we need the expansion to order $\vec{k}^{\,4}$.
The generic expansion in terms of $q^2$ is
\begin{align}
G_E(q) &=1 + {\langl r_E^2\rangl \over 6}  q^{\,2} + {\langl r_E^4\rangl \over 120}  q^{\,4} +\cdots \\
G_M (q) &=(1+\kappa) \left( 
1 + {\langl r_M^2\rangl \over 6}  q^{\,2} + {\langl r_M^4\rangl \over 120}  q^{\,4} +\cdots \right), \nonumber
\end{align}
where $\kappa$ is the anomalous magnetic moment, and $r_{E,M}$ is the  electric (magnetic) charge radius of the proton. This is the standard definition that gives charge conservation $G_E(0)=1$ and magnetic moment (or g-factor) $G_M(0)=1+\kappa$ at zero-momentum transfer.
The electric charge radii are defined as 
\beqs
\langl r_{E}^2\rangl &\equiv {6\over G_{E}(0)} {dG_{E} \over d  q^{\,2}} \Bigg |_{q^{\,2}=0},\\
\langl r_{E}^4\rangl &\equiv {120\over G_{E}(0)} {d^2G_{E} \over d  q^{\,4}} \Bigg |_{q^{\,2}=0},\\
\eeqs
and magnetic charge radii  similarly defined.
The Dirac form factors then take the forms, 
\begin{align}
& F_1 = {G_E + \tau G_M \over 1+\tau}  \nonumber \\
&=1-   \frac{1}{12}\left( {3\kappa \over m_p^2} - 2{\langl r_E^2\rangl} \right) q^{\,2}  \nonumber\\
&+ {5 \langl r_E^2\rangl + m_p^2 \langl r_E^4\rangl\over 120 m_p^2} q^{\,4}  \\
&- {m_p^2 \langl r_M^2\rangl +
   \kappa (3 + m_p^2 \langl r_M^2\rangl)\over 24 m_p^4} q^{\,4} \nonumber \\
&+\cdots, \nonumber
\end{align}
\begin{align}
& F_2 = {G_M - G_E \over 1+\tau} \nonumber  \\
&= \kappa- {1\over 12} \left(- {3\kappa \over m_p^2} + 2\langl r_E^2\rangl - 2(1+\kappa) \langl r_M^2\rangl \right)  q^{\,2} \nonumber \\
&- {  {5  \langl r_E^2\rangl} + m_p^2 \langl r_E^4\rangl \over 120 m_p^2} q^{\,4}  \\
&+ {{15\kappa+m_p^2(1+\kappa) ({5\langl r_M^2\rangl} + m_p^2 \langl  r_M^4\rangl)}\over 120 m_p^4} q^{\,4} \nonumber \\
&+\cdots, \nonumber
\end{align}
where 
$
\tau \equiv {-q^2 / (4m_p^2)}.
$

The discussion so far is general.
For electric polarizability, we need to work with the $\mu=\nu=0$ component, 
\begin{align}
& T_{00}^{Born} ={1\over 4m_p \vec{k}^{\,2}} \sum_{s_1,s_2} 
\left[
\bar{u}(\vec{0},s_2)
 \left(\gamma_0 F_1 - {i F_2\over 2m_p}\sigma_{03} k_z \right) \right. \nonumber \\
& \left. (\gamma_0 m_p+\gamma_3 k_z+m_p) 
\left(\gamma_0 F_1 + {i F_2\over 2m_p}\sigma_{03} k_z\right)
u(\vec{0},s_1)  \right. \nonumber \\
& \left. + \bar{u}(\vec{0},s_2) 
\left(\gamma_0 F_1 + {i F_2\over 2m_p} \sigma_{03} k_z \right)  \right. \label{eq:TB00} \\
& \left. (\gamma_0 m_p-\gamma_3 k_z+m_p) 
\left(\gamma_1 F_1 - {i F_2\over 2m_p}\sigma_{03} k_z\right)
u(\vec{0},s_1)
\right], \nonumber  
\end{align}
where $k_z$ refers to the spatial momentum in the z-direction.
Evaluating this expression, we find that the diagonal spin terms $s_1=s_2=1 \text{ or } 2$ give the same results whereas the off-diagonal terms vanish.
The final result is 
\beqs
&T_{00}^{Born}(\vec{k})  = {4m_p \over \vec{k}^{\,2}} - {4\over 3} \langl r_E^2\rangl m_p \\
&+\left[ -{\kappa^2\over 4m_p^3} + {m_p\over 45} \left(5\langl r_E^2\rangl^2 +3\langl r_E^4\rangl\right)  \right]  \vec{k}^{\,2} +\cdots.
\label{eq:proelas_elec_Born}
\eeqs
Including the contact interaction term,
 the full amplitude in the continuum takes the form,
\beq
T_{00}(\vec{k}) = T_{00}^{Born}(\vec{k}) +  \vec{k}^{\,2} {\alpha^p_E \over \alpha}.
\eeq

On the other hand, we consider the unpolarized four-point function of the proton in lattice regularization,
\begin{align}
&T_{\mu\nu}=i\ N_s a {1\over 2} \sum_{s_1,s_2} 
\int_{-\infty}^\infty dt \sum_{\vec{x}} e^{ik_2\cdot x} \\ 
&\langl p_2,s_2|\left[ T j^L_\mu(x) j^L_\nu(0) 
- \langl 0 |T j^L_\mu(x) j^L_\nu(0) |0\rangl \right] |p_1,s_1\rangl, \nonumber
\end{align}
where the VEV subtraction is included.
After inserting a complete set of intermediate states, 
\beq
\sum_{N,\vec{p}_N,s_N}  | (E_N,\vec{p}_N),s_N \rangl \langl (E_N,\vec{p}_N),s_N |=1,
\eeq
and specializing to the zero-momentum Breit frame, we have
\begin{align}
&T_{\mu\nu} =
N_s^2 \sum_{N,s_1,s_2,s_N}  {1 \over E_N-m_p } \times \label{eq:Tp} \\
&\langl (m_p,\vec{0}),s_2 | j^L_\mu | (E_N,\vec{q}),s_N)\rangl
\langl (E_N,\vec{q}),s_N) | j^L_\nu | (m_p,\vec{0}),s_1 \rangl \nonumber\\
&-
N_s^2 \sum_{N,s_1,s_2,s_N}  {1 \over E_N } \times \nonumber\\
&\langl 0 | j^L_\mu(0) | (E_N,\vec{q}),s_N)\rangl 
\langl (E_N,\vec{q}),s_N) | j^L_\nu(0) | 0 \rangl.
\nonumber
\end{align}
Due to the vector nature of the electromagnetic current, the only intermediate states that can contribute are spin-1/2 and spin-3/2 states.
We separate off the elastic part (N=proton term in the first sum),
\begin{align}
&T^{elas}_{\mu\nu} \equiv
N_s^2 \sum_{s_1,s_2,s_p}  {1 \over E_p-m_p } \times \label{eq:Telas1} \\
&\langl (m_p,\vec{0}),s_2 | j^L_\mu | (E_p,\vec{q}),s_p)\rangl
\langl (E_p,\vec{q}),s_p) | j^L_\nu | (m_p,\vec{0}),s_1 \rangl. \nonumber
\end{align}
The remaining inelastic part will be related to polarizabilities. 
The connection between the lattice and continuum matrix elements is 
\beq
\langl p',s' | j^L_\mu(0) | p,s \rangl
= {1\over N_s}
{(p',s' | j_\mu(0) |p,s ) \over \sqrt{2E_p 2E_{p'}} }.
\eeq
Using the continuum definition of form factors ($q=p'-p$),
\beq
( p',s' | j_\mu |p,s )=
\bar{u}(p',s') 
\left(\gamma_\mu F_1(q) + {iF_2(q)\over 2m_p}\sigma_{\mu\lambda} q^\lambda\right) 
u(p,s),
\eeq
 the elastic part can be written,
\beqs
T^{elas}_{\mu\nu} &=
 \sum_{s_1,s_2}  {1 \over 4m_p E_p(E_p-m_p) }  \\
&\bar{u}(\vec{0},s_2)  
\left(\gamma_\mu F_1 - {iF_2\over 2m_p}\sigma_{\mu\lambda} q^\lambda\right) 
 (\fsl{q}+m_p)  \\
&\left(\gamma_\nu F_1 + {iF_2\over 2m_p}\sigma_{\nu\lambda} q^\lambda\right)
u(\vec{0},s_1), 
 \label{eq:Telas2} 
\eeqs
where $q=(E_p,0,0,q_z)$ is the on-shell proton and we have used the spin sum, 
\beq
\sum_s u(p,s)\bar{u}(p,s) =\fsl{p}+m_p.
\eeq

For electric polarizability, we are interested in the  $\mu=\nu=0$ component of Eq.\eqref{eq:Telas2},
\beqs
&T^{elas}_{00} =
 \sum_{s_1,s_2}  {1 \over 4m_p E_p(E_p-m_p) }  \\
&\bar{u}(\vec{0},s_2)  
\left(\gamma_0 F_1 - {iF_2\over 2m_p}\sigma_{03} q_z\right) 
 (\gamma_0 E_p +\gamma_3 q_z+m_p)  \\
&\left(\gamma_0 F_1 + {iF_2\over 2m_p}\sigma_{03} q_z\right)
u(\vec{0},s_1).
 \label{eq:Telas3} 
\eeqs
It evaluates to order $\vec{q}^{\,2}$ as, 
\begin{align}
&T_{00}^{elas}(\vec{q})  = {4m_p \over \vec{q}^{\,2}}  - {4\over 3} \langl r_E^2\rangl m_p \\
&+ 
\left[{\langl r_E^2\rangl \over 3m_p}+ \frac{1}{4m_p^3}+\frac{m_p}{45}\left(5\langl r_E^2\rangl^2 +3\langl r_E^4\rangl\right)  \right]  \vec{q}^{\,2} +\cdots.
 \nonumber
  \label{eq:Telas4} 
\end{align}

Matching the lattice and continuum forms and subtracting off the elastic contribution, we have
\beq
T_{00}(\vec{q}) - T_{00}^{elas}(\vec{q}) =  T_{00}^{Born}(\vec{q}) - T_{00}^{elas}(\vec{q}) + \vec{q}^{\,2} {\alpha^p_E \over \alpha}. 
\eeq
Many terms cancel between $T_{00}^{Born}$ and $T_{00}^{elas}$, leaving the difference,
\beq
T_{00}(\vec{q}) - T_{00}^{elas}(\vec{q}) = \bigg[-\frac{\langl r_{E}^2\rangl}{3m_{p}} -{1+\kappa^2 \over 4 m_p^3} 
+ {\alpha^p_E \over \alpha} \bigg]  \vec{q}^{\,2},
\eeq
from which we arrive at a final formula for proton electric polarizability,
\beq
\alpha^p_E=  \alpha \left[  \frac{\langl r_{E}^2\rangl}{3m_{p}} +{1+\kappa^2 \over 4 m_p^3} 
+ {T_{00}(\vec{q}_1) - T_{00}^{elas}(\vec{q}_1) \over  \vec{q}_1^{\,2} } \right].
\label{eq:alpha_p}
\eeq
Here we emphasize that the expression must be evaluated using the smallest nonzero momentum  $\vec{q}_1$  on the lattice.
Compared to charged pion electric polarizability $\alpha^\pi_E$  in Eq.\eqref{eq:alpha_pion}, proton $\alpha^p_E$ has an extra term that has its magnetic moment and mass. In this sense, the proton's electric and magnetic properties are coupled. The $m_p$, $\langl r_{E}^2\rangl$, and $\kappa$ have to be measured at the same time as $T_{00}$ in order to extract $\alpha^p_E$.

%-----------------------
\subsection{Magnetic polarizability}

For the Compton amplitude in the continuum,
we start with the $\mu=\nu=1$ component of Eq.\eqref{eq:TB} (22 component gives the same result),
\begin{align}
&T_{11}^{Born} ={1\over 4m_p \vec{k}^{\,2}} \sum_{s_1,s_2} 
\left[
\bar{u}(\vec{0},s_2)
 \left(\gamma_1 F_1 - {iF_2\over 2m_p}\sigma_{13} k_z \right) \right. \nonumber \\
& \left. (\gamma_0 m_p+\gamma_3 k_z+m_p) 
\left(\gamma_1 F_1 + {iF_2\over 2m_p}\sigma_{13} k_z\right)
u(\vec{0},s_1)  \right. \nonumber \\
& \left. + \bar{u}(\vec{0},s_2) 
\left(\gamma_1 F_1 + {iF_2\over 2m_p}\sigma_{13} k_z \right)  \right. \label{eq:TB11} \\
& \left. (\gamma_0 m_p-\gamma_3 k_z+m_p) 
\left(\gamma_1 F_1 - {iF_2\over 2m_p}\sigma_{13} k_z\right)
u(\vec{0},s_1)
\right]. \nonumber  
\end{align}
 It evaluates to
\beqs
&T_{11}^{Born}(\vec{k})  = {\kappa(2+\kappa) \over m_p} \\
&+ \vec{k}^{\,2}  
\left( - {\kappa \over 2m_p^3} +  {\langl r_E^2\rangl \over 3m_p}
 -  {\langl r_M^2\rangl \over 3m_p} (1+\kappa)^2 \right).
\eeqs
Including the contact interaction term, the full amplitude in the continuum becomes
\beq
T_{11}(\vec{k}) = T_{11}^{Born}(\vec{k}) +  \vec{k}^{\,2} {\beta^p_M \over \alpha}.
\label{eq:T11_cont}
\eeq

On the lattice, we start with the $\mu=\nu=1$ component of Eq.\eqref{eq:Telas2}, 
\beqs
&T^{elas}_{11} =
 \sum_{s_1,s_2}  {1 \over 4m_p E_p(E_p-m_p) }  \\
&\bar{u}(\vec{0},s_2)  
\left(\gamma_1 F_1 + {iF_2\over 2m_p}\big[-\sigma_{13} q_z+\sigma_{10} (m_p-E_p)\big]\right) \\ &
 (\gamma_0 E_p+\gamma_3 q_z+m_p)  \\
&\left(\gamma_1 F_1 + {iF_2\over 2m_p}\big[\sigma_{13} q_z+\sigma_{10} (E_p-m_p)\big]\right)
u(\vec{0},s_1).
 \label{eq:Telas3_p} 
\eeqs
It evaluates to
\beq
T_{11}^{elas}(\vec{q})  =(1+\kappa)^2  \bigg[{1 \over m_p} 
- \vec{q}^{\,2}  
\left( {1 \over 2m_p^3} +  {\langl r_M^2\rangl \over3 m_p}\right) \bigg] .
\label{eq:T11_elas_p}
\eeq
We see that unlike a charged pion, there is an elastic contribution for the proton magnetic case. 

The inelastic 11 component in Eq.\eqref{eq:Tp} can be formally characterized as a constant plus a linear term in $\vec{q}^{\,2}$, 
\beq
T_{11}^{inel}(\vec{q}) \equiv T_{11}^{inel}(\vec{0}) +  \vec{q}^{\,2} K_{11},
\label{eq:inelastic_lat_p}
\eeq
with $T_{11}^{inel}(\vec{0})$ and $K_{11}$ to be matched with physical parameters.
The difference between the Born term in the continuum and the elastic term on the lattice is 
\beq
T_{11}^{Born}-T_{11}^{elas} = 
-{1 \over m_p} + \vec{q}^{\,2}  
\left(  {1+\kappa+\kappa^2 \over 2m_p^3} +  {\langl r_E^2\rangl \over 3m_p} \right),
\label{eq:T11diff}
\eeq
where the $\kappa$ terms in the zero-momentum part cancel, as well as the magnetic charge radius terms in the $\vec{q}^{\,2}$ part. 

By matching the full $T_{11}$ in the continuum and on the lattice, we have,
\beq
T_{11}^{elas}(\vec{q}) + T_{11}^{inel}(\vec{0}) +  \vec{q}^{\,2} K_{11} =T_{11}^{Born}(\vec{q}) +  \vec{q}^{\,2} {\beta^p_M\over \alpha}.
\eeq
Using Eq.\eqref{eq:T11diff}, we obtain two relations,
\begin{align}
 T_{11}^{inel}(\vec{0}) &= -{1\over m_p}, \label{eq:T11_proton}\\
K_{11} &= {\langl r_E^2\rangl \over 3m_p} + {1+\kappa+\kappa^2 \over 2m_p^3}   + {\beta^p_M \over \alpha }.
 \label{eq:K11_proton}
\end{align}
We see the same sum rule in the first relation as Eq.\eqref{eq:sr} for charged pion. The second relation produces an expression for proton magnetic polarizability on the lattice, 
\beq
\beta^p_M  = \alpha \left[-  {\langl r_E^2\rangl \over 3m_p} - {1+\kappa+\kappa^2 \over 2m_p^3} 
+ { T^{inel}_{11}(\vec{q}_1) - T^{inel}_{11}(\vec{0}) \over \vec{q}_1^{\,2}} \right],
\eeq
where we have used Eq.\eqref{eq:inelastic_lat_p} for $K_{11}$.

It turns out there is no elastic part to the zero-momentum amplitude $T_{11}(\vec{0})$.
There is a subtlety here. If we do the analytic time integral first, then set $\vec{q}=0$, we get $T_{11}^{elas}(\vec{0})=(1+\kappa)^2/m_p$ from Eq.\eqref{eq:T11_elas_p}.
However, if we first set $\vec{q}=0$, the integrand itself vanishes, so $T_{11}^{clas}(\vec{0})=0$.
This is the way it is done on the lattice in a numerical sense as we see in Eq.\eqref{eq:Q11elas}.
So we can drop the reference to the inelastic part $T^{inel}_{11}(\vec{0})\to T_{11}(\vec{0})$.
Using the full amplitude $T_{11}$ defined in Eq.\eqref{eq:Tp}, we write the final lattice formula for proton magnetic polarizability as, 
\beqs
\beta^p_M & = \alpha \big[-  {\langl r_E^2\rangl \over 3m_p} - {1+\kappa+\kappa^2 \over 2m_p^3} \\&
+ { T_{11}(\vec{q}_1) - T^{elas}_{11}(\vec{q}_1) - T_{11}(\vec{0}) \over \vec{q}_1^{\,2}} \big].
 \label{eq:beta_p}
\eeqs
Compared to charged pion magnetic polarizability $\beta^\pi_M$  in Eq.\eqref{eq:beta_pion}, proton $\beta^p_M$ has two extra terms: 
a mass contribution and an elastic contribution. 
Both terms, along with the $r_E$ term, 
must be measured at the same time as $T_{11}$ in order to extract $\beta^p_M$.

%%%%%%%%%%%%%%%%%%%%%%%%%%%%%
\section{Lattice measurement}
\label{sec:lat} 
\begin{figure}[h]
\includegraphics[scale=0.55]{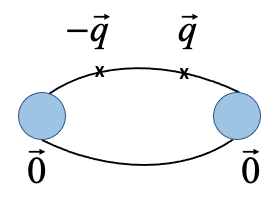}
\caption{Zero-momentum Breit frame in Eq.\eqref{eq:Breit2} used in extracting charged pion polarizabilities from four-point functions on the lattice (for proton imagine three quark lines). Time flows from right to left and the four-momentum conservation is expressed as $p_2 =q_2+q_1+p_1$.}
\label{fig:diagram-4pt2}
\end{figure}

Having obtained polarizability formulas in Eq.\eqref{eq:alpha_pion} and Eq.\eqref{eq:beta_pion} for a charged pion,
and Eq.\eqref{eq:alpha_p} and Eq.\eqref{eq:beta_p} for a proton,
we  now discuss how to measure them in lattice QCD.
First, we switch from Minkowski to Euclidean time on the lattice. 
Next, we need to match the kinematics used in deriving the expressions, {\it i.e.}, with hadrons at rest and photons having spacelike momentum in the z-direction~\footnote{For general discussion, we use 
$h$ to represent either a charged pion or proton.},
\beqs
p_1 &=(m_h,0,0,0),\\
q_1 &=(0,0,0,q_z),\;
q_2 =(0, 0,0,-q_z), \; q_z \ll m_h,\\
p_2 &=q_2+q_1+p_1=(m_h,0,0,0),
\label{eq:Breit2}
\eeqs
as illustrated in Fig~\ref{fig:diagram-4pt2}.
It is the same kinematics as in Eq.~\eqref{eq:Breit1} but expressed differently to match what is being done on the lattice. One may think of Fig~\ref{fig:diagram-4pt2} as having ``internal" photons, whereas Fig~\ref{fig:diagram-4pt1} as having ``external" photons.

We construct the four-point current-current correlation function,
\begin{align}
&P_{\mu\nu}(\vec{x}_2,\vec{x}_1,t_2,t_1) \equiv   \label{eq:P1} \\  & 
 {%\smashoperator[r]
 {\sum_{\vec{x}_3, \vec{x}_0}} \langl 0 | \psi^\dagger (x_3) :j^L_\mu(x_2) j^L_\nu(x_1):  \psi (x_0) |0\rangl
\over 
\sum_{\vec{x}_3,\vec{x}_0} \langl 0 | \psi^\dagger (x_3) \psi (x_0) |0\rangl \nonumber
},
\end{align}
where the two-point function is for normalization, $\psi$ the interpolating field of the hadron, and normal ordering is used to include the VEV contribution. In the case of a proton, sum over final spin and average over initial spin are assumed for unpolarized measurement.
The spatial sums over $\vec{x}_3$ and $\vec{x}_0$ project to zero momentum at the sources which are located at fixed times $t_3$ and $t_0$. Time flows from right to left $t_3>t_{1,2}>t_0$, and $t_{1,2}$ indicates the two possibilities of time ordering.
Zero-momentum sources can be realized by wall sources without gauge fixing~\cite{WallSource1993,Fu_2012}. 
The formula is now in discrete Euclidean spacetime but we keep the Euclidean time axis continuous.
Here, the current is the lattice version of Eq.\eqref{eq:current}, preferably the conserved current positioned as symmetrically as possible about the sources.
When the times are well separated (defined by the time limits $t_3\gg t_{1,2} \gg t_0$) the correlator is dominated by
the ground state, 
\fontsize{9}{9}
\beqs
&P_{\mu\nu}(\vec{x}_2,\vec{x}_1,t_3,t_2,t_1,t_0)  \to \langl \pi(\vec{0}) | : j^{L}_\mu(x_2) j^{L}_\nu(x_1): |  \pi(\vec{0})\rangl \\
&=\langl \pi(\vec{0}) | T j^{L}_\mu(x_2) j^{L}_\nu(x_1) |  \pi(\vec{0})\rangl  
 -  \langl 0 | T j^{L}_\mu(x_2) j^{L}_\nu(x_1) | 0\rangl.
 \eeqs
To implement the special kinematics, we consider the Fourier transform (suppressing $t_3$ and $t_0$ for clarity)
\fontsize{9}{9}
\beq
Q_{\mu\nu}(\vec{q},t_2,t_1)  \equiv N_s\sum_{\vec{x}_2,\vec{x}_1} e^{-i\vec{q}\cdot \vec{x}_2} e^{i\vec{q}\cdot \vec{x}_1} 
P_{\mu\nu}(\vec{x}_2,\vec{x}_1,t_3,t_2,t_1,t_0),
\label{eq:Qmn}
\eeq
where $\vec{q}$ is lattice momentum injected at $\vec{x}_1$ and taken out at $\vec{x}_2$.
The need for Fourier transform is natural in the sense that  the polarizability formulas are derived in momentum space.
In this work we only consider the diagonal components ($\mu=\nu$) of $Q_{\mu\nu}(\vec{q},t_2,t_1) $.
Assuming the time separation $t=t_2-t_1>0 $ and inserting a complete set of intermediate states,  
the expression in the same time limits develops the time dependence,
\normalsize
\begin{align}
Q_{\mu\mu}(\vec{q},t) & =N^2_s \sum_{n} 
| \langl h(\vec{0}) | j^L_\mu(0) | n(\vec{q}) \rangl|^2  e^{-a(E_n-m_h) t}  \nonumber \\ 
&- N^2_s \sum_{n}  |\langl 0 | j^L_\mu(0) | n(\vec{q}) \rangl |^2  e^{-aE_n t}.
\label{eq:Q1} 
\end{align}
%
%where the $\vec{r}$ sum collapsed to give $\vec{q}$ to the intermediate states $n(\vec{q})$.
The elastic contribution ($n=h$ in the first sum) in the expression can be separately defined,
\beq
Q^{elas}_{\mu\mu}(\vec{q},t)   \equiv N^2_s 
|\langl h(\vec{0}) | j^L_\mu(0) | h(\vec{q}) \rangl |^2   e^{-a(E_h-m_h) t}.
\label{eq:Q2} 
\eeq
We see that the elastic piece in the four-point function has information on the form factors of the hadron through the amplitude and can be isolated at large time separations of the currents.

Charged pion electric polarizability in Eq.\eqref{eq:alpha_pion} is measured on the lattice by 
\beq
\alpha^{\pi}_E=\alpha \left\{\frac{\langl r^2\rangl}{3m_{\pi}}+{2 a \over \vec{q}_1^{\,2}} \int_{0}^\infty d t \left[Q_{00}(\vec{q}_1,t) -Q^{elas}_{00}(\vec{q}_1,t) \right]\right\}.
\label{eq:alpha_pion_lat}
\eeq
A few comments are in order. 
First, this work corrects two formulas in Ref.~\cite{Wilcox:1996vx}: Eq.(17) and  Eq.(81) here  should replace Eq.(26) and Eq.(45) there, respectively.
Second, 
the expression contains the electric charge radius squared $\langl r^2\rangl$ contribution which has to be added to the time integral. 
Fortunately, the four-point function $Q_{00}(\vec{q}_1,t) $ already contains information on the form factor in its elastic limit~\cite{Wilcox_1992,Andersen:1996qb},
\beq 
Q^{elas}_{00}(\vec{q}_1,t)\xrightarrow[t\gg 1]{} \big(1-{\langl r\rangl^2 \vec{q}_1^{\,2} \over 3}\big)  e^{-a(E_\pi-m_\pi) t}.
\eeq
It is just a matter of performing a separate analysis on the same correlators at large time separation of the two currents to extract the charge radius.
Third, $\alpha^\pi_E$ has the physical unit of $a^3$ (fm$^3$) since $1/\vec{q}_1^{\,2}$ scales like $a^2$, and  $Q_{00}$ and $t$ are dimensionless by definition. 
Fourth, the contribution from the time integral to $\alpha^\pi_E$ is proportional to the difference in the areas under the two curves. 
It is this difference that determines the sign of $\alpha^\pi_E$.
On a finite lattice, the time integrals do not really extend to $\infty$ but are limited to the available 
time slices between the two insertions. In practice, one should check if the largest time separation is enough to establish the elastic limit.
Fifth, Eq.\eqref{eq:alpha_pion_lat} can be regarded as the numerical derivative of the time integral with respect to 
$\vec{q}_1^{\,2}$, evaluated at $\vec{q}_1^{\,2}=0$, with error on the order of $\vec{q}_1^{\,2}$. So the smaller the momentum is, the better the prediction is. This is the reason the smallest nonzero momentum $\vec{q}_1$ on the lattice should be used. 
Equivalent directions for $\vec{q}_1$ can be used to improve the signal-to-noise ratio.
Sixth, we emphasize the importance of working in momentum space to extract polarizabilities from four-point functions. In position space, once would deal with the quantity $R(t)=\sum_{\vec{r}} \vec{r}^{\,2} P_{\mu\nu} (\vec{r},t)$ with $\vec{r}=\vec{x}_2-\vec{x}_1$, which does not project to small momentum on the lattice~\cite{Wilcox_2002} and can lead to erroneous results.

Charged pion magnetic polarizability in Eq.\eqref{eq:beta_pion} is measured on the lattice by
\beq
\beta^{\pi}_M =\alpha \left\{- { \langl r^2\rangl \over 3m_\pi}
 +{2a\over  \vec{q}_1^{\,2}} \int_{0}^\infty d t \left[Q_{11}(\vec{q}_1,t) -Q_{11}(\vec{0},t)\right] \right\}, 
\label{eq:beta_pion_lat}
\eeq
where $Q_{11}(\vec{q}_1,t)$ is the 11 component of Eq.\eqref{eq:Q1}. 
%in the valence approximation, defined as  excluding the quark loop contributions, but including the VEV contributions. 
Unlike the electric case where the elastic contribution is subtracted in the time integral, the magnetic case has the zero-momentum inelastic contribution subtracted.
% The complication can be regarded as a blessing in disguise: it offers an alternative to the standard three-point function method for the form factors.
The sign of $\beta^\pi_M$ is dictated by the relative magnitudes of the charge radius terms and the time integral term which can be both positive or negative (see discussion below).

We now turn to the proton. The electric polarizability in Eq.\eqref{eq:alpha_p} can be measured on the lattice by
\begin{align}
&\alpha^{p}_E =\alpha  \left\{ 
\frac{\langl r_{E}^2\rangl}{3m_{p}}+{1+\kappa^2 \over 4 m_p^3} \right. \label{eq:alpha_p_lat} \\
& \left. +{2a\over \vec{q}_1^{\,2}} \int_{0}^\infty d t \left[Q_{00}(\vec{q}_1,t) -Q^{elas}_{00}(\vec{q}_1,t)\right] \right\},
 \nonumber
\end{align}
and the magnetic polarizability in Eq.\eqref{eq:beta_p} by
\begin{align}
&\beta^{p}_M =\alpha  \bigg\{-  {\langl r_E^2\rangl \over 3m_p} - {1+\kappa+\kappa^2 \over 2m_p^3}  \label{eq:beta_p_lat} \\
&  +{2a\over  \vec{q}_1^{\,2}} \int_{0}^\infty d t \left[Q_{11}(\vec{q}_1,t) -Q^{elas}_{11}(\vec{q}_1,t) -Q_{11}(\vec{0},t)\right] \big\}. 
 \nonumber
\end{align}
Most of the above-mentioned arguments for a charged pion apply also to the proton. 
The difference is they additionally involve the proton mass ($m_p$) and its anomalous magnetic moment ($\kappa$).  Both need to be measured along with the time integral on the same lattice.
The mass can be readily obtained from the two-point function which is already used in Eq.\eqref{eq:P1} for normalization. An excellent signal is expected for the mass measurement since well-separated zero-momentum sources are used. 
Although only the $Q_{00}$ component is needed for the time integral for $\alpha^p_E$, the elastic part of the $Q_{11}$ component  is required for the anomalous magnetic moment term in $\alpha^p_E$.
It can be extracted from the large-time behavior of,
\beq
Q^{elas}_{11}(\vec{q}_1,t)\xrightarrow[t\gg 1]{} {(1+\kappa)^2 \over 4 m_p^2} \vec{q}_1^{\,2} e^{-a(E_p-m_p) t}.
\label{eq:Q11elas}
\eeq
Since the $Q_{11}$ component is needed anyway in the calculation of $\beta^p_M$, the two measurements complement each other. The same is true of the charge radius term in $\beta^p_M$ which can be accessed through the unpolarized elastic part of $Q_{00}$,
\beq
Q^{elas}_{00}(\vec{q}_1,t)\xrightarrow[t\gg 1]{} 
\bigg[1-\vec{q}_1^{\,2}\big({ 1 \over 4 m_p^2}+{\langl r_E^2\rangl\over 3}\big) \bigg]e^{-a(E_p-m_p) t}.
\eeq
The close coupling between the electric and magnetic suggests that it is most efficient to measure the two polarizabilities together, with associated mass, charge radius, and magnetic moment in the same simulation. In practice, this should be done on a configuration by configuration basis to maintain correlations.

\begin{figure}
\includegraphics[scale=0.47]{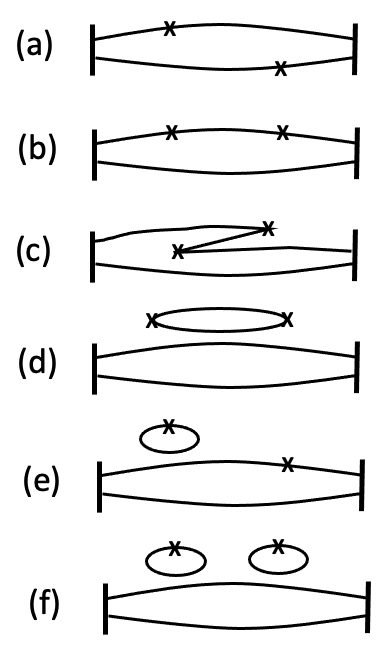}
\caption{Quark-line diagrams of a four-point function contributing to polarizabilities of a meson: (a) different flavor, (b) same flavor, (c) same flavor Z-graph, (d) single-flavor double-current loop, (e) single disconnected loop, (f) double disconnected loops.  In each diagram, flavor permutations are assumed as well as gluon lines that connect the quark lines. The zero-momentum hadron interpolating fields are represented by vertical bars (wall sources). For a baryon, imagine three quark lines between the sources instead of two.}
\label{fig:diagram-4pt3}
\end{figure}

Next, we discuss how to evaluate Eq.\eqref{eq:P1} and its Fourier transform Eq.\eqref{eq:Qmn} at the quark level.
Wick contractions of quark-antiquark pairs in the unsubtracted part lead to topologically 
distinct quark-line diagrams shown in Fig.~\ref{fig:diagram-4pt3}.
Diagrams (a), (b), and (c) are connected. Diagram (d) has a loop that is disconnected from the hadron but connected between the two currents. Diagram (e) has one disconnected loop (also known as all-to-all propagator), and diagram (f) has two such loops. 
Furthermore, diagrams (d), (e), and (f) must have associated VEV subtracted.
However, if conserved lattice current density is used, there is no need for subtraction in diagram (e)
since the VEV vanishes in the configuration average~\cite{Draper:1988bp}.

According to Eq.\eqref{eq:current}, the full hadron polarizabilities can be broken down to contributions from various quark flavor current-current correlations. Assuming isospin symmetry in $u$ and $d$ quarks, we have
\beqs
\alpha_E^{\pi}&={5\over 9} \alpha^{uu}_E +{4\over 9} \alpha^{u\bar{d}}_E, \\
\beta_M^{\pi}&={5\over 9} \beta^{uu}_M + {4\over 9} \beta^{u\bar{d}}_M,
\label{eq:ud_pion}
\eeqs
for a charged pion, and 
\beqs
\alpha_E^{p}&={4\over 9} \alpha^{uu}_E + {1\over 9} \alpha^{dd}_E  -{4\over 9} \alpha^{ud}_E, \\
\beta_M^{p}&={4\over 9} \beta^{uu}_M + {1\over 9} \beta^{dd}_M  -{4\over 9} \beta^{ud}_M,
\label{eq:ud_p}
\eeqs
for a proton.
Specifically, the quark flavor labels $uu$, $dd$, $ud$, and $u\bar{d}$ refer to contributions in Eq.\eqref{eq:Q1} and Eq.\eqref{eq:Q2} without the charge factors which have been pulled out in Eq.\eqref{eq:ud_pion} and Eq.\eqref{eq:ud_p}.
Under the same isospin symmetry $\pi^{+}$ and $\pi^{-}$ have identical polarizabilities.
These decompositions, together with the diagrams in Fig.~\ref{fig:diagram-4pt3}, provide a physical picture of how polarizabilities arise from quark-gluon dynamics of QCD.

Computationally, the different flavor diagram (a) for a charged pion is the easiest to simulate. The pion sources can be fixed at the two time ends and the quark lines are ``sewn" together.
For the proton, there are more flavor permutations to consider since a proton has two $u$ quark lines and one $d$ quark line. For example, one current coupling to the first $u$ quark and the other to the second $u$ quark is considered a ``different flavor" diagram. This diagram requires both $x_1$ and $x_2$ to vary.
At each relative time separation, the Fourier transform involves double spatial sums over the entire lattice that have the form
\beq
f(\vec{q})=\sum_{\vec{r}} e^{-i\vec{q}\cdot \vec{r}} \sum_{\vec{x}} f_1(\vec{r}+\vec{x}) f_2(\vec{x}).
\eeq
The $\vec{x}$ sum can be sped up by using fast Fourier transform algorithms~\cite{Andersen:1996qb}. 

In diagrams (b) and (c), the currents couple to the same quark line, and a sequential source technique (SST) is necessary. One approach is to fix the time position of the two hadron fields and the initial source. In Ref.~\cite{Wilcox:1993uq}, for example, both the final pion and the initial current source were put in as sources. For diagram (b), the final pion and the initial current are put in as separate SST sources on the initial pion quark line. For the Z-graph, both the final pion and the initial source are put in as two consecutive SST sources on the initial quark lines. In order to enhance the signal, a technique called Fourier reinforcement~\cite{Wilcox:1993uq} can be applied to reduce the associated statistical errors. This technique smears over the initial current source in two spatial directions, leaving one direction unrestricted. This reinforces the signal but restricts the Fourier transform to the remaining direction.

In diagram (d), both currents couple to the same sea quark loop. 
The loop originates together with diagram (f) as the disconnected part of the current-current correlation but can be evaluated relatively easily using two-point quark propagators. The only complication is that it has a VEV that needs to be subtracted before the loop is correlated with the hadron propagator. This diagram is expected to produce the strongest signal from the sea quarks.  It will give us the first glimpse into sea-quark contributions to polarizabilities and form factors from four-point functions. 

Diagram (e) comes from the connected part of the current-current correlation with one current still connected to the hadron propagator and the other disconnected. In diagram (f), not only are the two currents disconnected from each other, but their loops are also disconnected from the hadron propagator.  Diagram (f) is expected to produce the weakest signal, and it is also the most challenging to simulate. 
Diagram (e) does not have a VEV in the configuration average but diagram (f) does. 
All-to-all quark propagators have been encountered in other studies of hadron structure. One can draw on lattice noise methods such as Refs.~\cite{BARAL201964,Romero_2020,Liu:2017man,Morningstar_2011} to extract a signal.

%%%%%%%%%%%%%%%%%%%%%%%%%%%
\section{\label{sec:con} Conclusion}
Computing polarizability of charged hadrons has been a challenge for lattice QCD due to the acceleration and Landau levels in the background field method. 
In this work, we lay out a program for the use of four-point correlation functions as an alternative, 
by revitalizing an earlier study on electric polarizability of charged pions and expanding the formalism to include magnetic polarizability and the proton.
The approach bears a close resemblance to the Compton scattering process 
with a transparent physical picture and conceptual clarity.
%The trade-off is the increased computational demand related to four-point functions.

We detailed how to construct lattice correlators and methods to isolate the polarizabilities. In the case of $\alpha^\pi_E$, there are no technical hurdles on the connected contributions,  as demonstrated by the signals in Refs.~\cite{Wilcox:1996vx,Wilcox:1993uq} even with relatively primitive lattices by today's standards. In the case of $\beta^\pi_M$, the charge radius and mass need to be measured on the same lattice in addition to the time integral, but the information is already present in the elastic limit for $\alpha^\pi_E$. 
This is an added advantage of the current approach: the elastic limit of four-point correlation functions also offers an alternative to form factors from the traditional three-point functions. 
For the proton, $\alpha^p_E$ and $\beta^p_M$ are more complicated; both have extra terms that involve mass, charge radius, or magnetic moment. But $\alpha^p_E$ and $\beta^p_M$ complement each other; the four-point function for $\alpha^p_E$ has information on the extra term(s) in the four-point function for $\beta^p_M$  and vice versa. For this reason, it is best to simulate $\alpha^h_E$ and $\beta^h_M$ together, completing all relevant measurements (mass from two-point function, charge radius and magnetic moment from elastic limit of four-point functions $Q_{00}$ and  $Q_{11}$, and time integrals) on a configuration before moving to the next one to maintain correlations in the parameters.

Here, we want to point out an important issue in the calculation of polarizabilities in lattice QCD, 
that is, sea-quark contributions. In the background field method, charging the sea quarks is a systematic uncertainty to be removed in almost all the existing calculations. The challenge lies in the fact that 
generating a separate Monte Carlo ensemble to compute the correlator in the presence of a background field would ruin the correlations relied upon for extracting a small mass shift. 
As a workaround, perturbative reweighting has been proposed as a method of creating two ensembles which have different sea-quark actions yet are correlated~\cite{Freeman:2014kka,Freeman:2013eta}. Such calculations are expensive and fermion action dependent.
In the four-point function approach, on the other hand, sea-quark charging is avoided. Sea-quark effects are automatically included by the self-contraction of current-coupled quark loops in diagrams (d), (e), and (f) in Fig.~\ref{fig:diagram-4pt3}. One can work with any existing dynamical configurations without modification.
Moreover, the strangeness contribution can be straightforwardly studied by adding a $s$-quark component to the current in Eq.\eqref{eq:current}.  

Finally, four-point function techniques are also useful for hadron structure function calculations leading to parton distribution functions. The same Compton meson and (unpolarized) baryon quark-line diagrams are evaluated, except now at high momentum transfer. For example, Ref.~\cite{Liang:2019frk} used a nonzero momentum Breit frame to evaluate the Fourier transformed $j_0(\vec{x},t)j_0(0)$ and $j_1(\vec{x},t)j_1(0)$ proton correlation functions, which are the same ones necessary for electric and magnetic polarizability. The key to this evaluation is the implementation of the inverse Laplace transform~\cite{Wilcox:1992xe},
such as the Bayesian reconstruction method employed in Ref.~\cite{Liang:2019frk}. Using this technique, useful comparisons on proposed continuum forms can be examined.

With the availability of state-of-the-art dynamical ensembles, more powerful computers,  and efficient algorithms, we believe the time has come to directly tackle the four-point functions to extract polarizabilities in lattice QCD simulations.

\vspace*{5mm}
\begin{acknowledgments}
We thank Keh-Fei Liu and Andrei Alexandru for reading the manuscript and helpful discussions as well as Xuan-He Wang and Yang Fu for catching an error in the intermediate steps for proton electric polarizability which did not affect the final results. W.W. would like to acknowledge a Baylor University Arts and Sciences Research Leave. 
This work was supported in part by DOE Grant~No.~DE-FG02-95ER40907. 
\end{acknowledgments}

%%%%%%%%%%%%%%%%%%%%%%%%%%%%
%\clearpage
\bibliography{x4ptfun}

\begin{thebibliography}{53}%
\makeatletter
\providecommand \@ifxundefined [1]{%
 \@ifx{#1\undefined}
}%
\providecommand \@ifnum [1]{%
 \ifnum #1\expandafter \@firstoftwo
 \else \expandafter \@secondoftwo
 \fi
}%
\providecommand \@ifx [1]{%
 \ifx #1\expandafter \@firstoftwo
 \else \expandafter \@secondoftwo
 \fi
}%
\providecommand \natexlab [1]{#1}%
\providecommand \enquote  [1]{``#1''}%
\providecommand \bibnamefont  [1]{#1}%
\providecommand \bibfnamefont [1]{#1}%
\providecommand \citenamefont [1]{#1}%
\providecommand \href@noop [0]{\@secondoftwo}%
\providecommand \href [0]{\begingroup \@sanitize@url \@href}%
\providecommand \@href[1]{\@@startlink{#1}\@@href}%
\providecommand \@@href[1]{\endgroup#1\@@endlink}%
\providecommand \@sanitize@url [0]{\catcode `\\12\catcode `\$12\catcode
  `\&12\catcode `\#12\catcode `\^12\catcode `\_12\catcode `\%12\relax}%
\providecommand \@@startlink[1]{}%
\providecommand \@@endlink[0]{}%
\providecommand \url  [0]{\begingroup\@sanitize@url \@url }%
\providecommand \@url [1]{\endgroup\@href {#1}{\urlprefix }}%
\providecommand \urlprefix  [0]{URL }%
\providecommand \Eprint [0]{\href }%
\providecommand \doibase [0]{http://dx.doi.org/}%
\providecommand \selectlanguage [0]{\@gobble}%
\providecommand \bibinfo  [0]{\@secondoftwo}%
\providecommand \bibfield  [0]{\@secondoftwo}%
\providecommand \translation [1]{[#1]}%
\providecommand \BibitemOpen [0]{}%
\providecommand \bibitemStop [0]{}%
\providecommand \bibitemNoStop [0]{.\EOS\space}%
\providecommand \EOS [0]{\spacefactor3000\relax}%
\providecommand \BibitemShut  [1]{\csname bibitem#1\endcsname}%
\let\auto@bib@innerbib\@empty
%</preamble>
\bibitem [{\citenamefont {L'vov}(1993)}]{Lvov:1993fp}%
  \BibitemOpen
  \bibfield  {author} {\bibinfo {author} {\bibfnamefont {A.~I.}\ \bibnamefont
  {L'vov}},\ }\bibfield  {title} {\enquote {\bibinfo {title} {{Theoretical
  aspects of the polarizability of the nucleon}},}\ }\href {\doibase
  10.1142/S0217751X93002095} {\bibfield  {journal} {\bibinfo  {journal} {Int.
  J. Mod. Phys. A}\ }\textbf {\bibinfo {volume} {8}},\ \bibinfo {pages}
  {5267--5303} (\bibinfo {year} {1993})}\BibitemShut {NoStop}%
\bibitem [{\citenamefont {L'vov}\ \emph {et~al.}(2001)\citenamefont {L'vov},
  \citenamefont {Scherer}, \citenamefont {Pasquini}, \citenamefont {Unkmeir},\
  and\ \citenamefont {Drechsel}}]{Lvov_2001}%
  \BibitemOpen
  \bibfield  {author} {\bibinfo {author} {\bibfnamefont {A.~I.}\ \bibnamefont
  {L'vov}}, \bibinfo {author} {\bibfnamefont {S.}~\bibnamefont {Scherer}},
  \bibinfo {author} {\bibfnamefont {B.}~\bibnamefont {Pasquini}}, \bibinfo
  {author} {\bibfnamefont {C.}~\bibnamefont {Unkmeir}}, \ and\ \bibinfo
  {author} {\bibfnamefont {D.}~\bibnamefont {Drechsel}},\ }\bibfield  {title}
  {\enquote {\bibinfo {title} {Generalized dipole polarizabilities and the
  spatial structure of hadrons},}\ }\href {\doibase 10.1103/PhysRevC.64.015203}
  {\bibfield  {journal} {\bibinfo  {journal} {Phys. Rev. C}\ }\textbf {\bibinfo
  {volume} {64}},\ \bibinfo {pages} {015203} (\bibinfo {year}
  {2001})}\BibitemShut {NoStop}%
\bibitem [{\citenamefont {Moinester}\ and\ \citenamefont
  {Scherer}(2019)}]{Moinester:2019sew}%
  \BibitemOpen
  \bibfield  {author} {\bibinfo {author} {\bibfnamefont {Murray}\ \bibnamefont
  {Moinester}}\ and\ \bibinfo {author} {\bibfnamefont {Stefan}\ \bibnamefont
  {Scherer}},\ }\bibfield  {title} {\enquote {\bibinfo {title} {{Compton
  Scattering off Pions and Electromagnetic Polarizabilities}},}\ }\href
  {\doibase 10.1142/S0217751X19300084} {\bibfield  {journal} {\bibinfo
  {journal} {Int. J. Mod. Phys. A}\ }\textbf {\bibinfo {volume} {34}},\
  \bibinfo {pages} {1930008} (\bibinfo {year} {2019})},\ \Eprint
  {http://arxiv.org/abs/1905.05640} {arXiv:1905.05640 [hep-ph]} \BibitemShut
  {NoStop}%
\bibitem [{\citenamefont {Lensky}\ and\ \citenamefont
  {Pascalutsa}(2010)}]{Lensky:2009uv}%
  \BibitemOpen
  \bibfield  {author} {\bibinfo {author} {\bibfnamefont {Vadim}\ \bibnamefont
  {Lensky}}\ and\ \bibinfo {author} {\bibfnamefont {Vladimir}\ \bibnamefont
  {Pascalutsa}},\ }\bibfield  {title} {\enquote {\bibinfo {title} {{Predictive
  powers of chiral perturbation theory in Compton scattering off protons}},}\
  }\href {\doibase 10.1140/epjc/s10052-009-1183-z} {\bibfield  {journal}
  {\bibinfo  {journal} {Eur. Phys. J. C}\ }\textbf {\bibinfo {volume} {65}},\
  \bibinfo {pages} {195--209} (\bibinfo {year} {2010})},\ \Eprint
  {http://arxiv.org/abs/0907.0451} {arXiv:0907.0451 [hep-ph]} \BibitemShut
  {NoStop}%
\bibitem [{\citenamefont {Hagelstein}(2020)}]{Hagelstein:2020vog}%
  \BibitemOpen
  \bibfield  {author} {\bibinfo {author} {\bibfnamefont {Franziska}\
  \bibnamefont {Hagelstein}},\ }\bibfield  {title} {\enquote {\bibinfo {title}
  {{Nucleon Polarizabilities and Compton Scattering as a Playground for Chiral
  Perturbation Theory}},}\ }\href {\doibase 10.3390/sym12091407} {\bibfield
  {journal} {\bibinfo  {journal} {Symmetry}\ }\textbf {\bibinfo {volume}
  {12}},\ \bibinfo {pages} {1407} (\bibinfo {year} {2020})},\ \Eprint
  {http://arxiv.org/abs/2006.16124} {arXiv:2006.16124 [nucl-th]} \BibitemShut
  {NoStop}%
\bibitem [{\citenamefont {McGovern}\ \emph {et~al.}(2013)\citenamefont
  {McGovern}, \citenamefont {Phillips},\ and\ \citenamefont
  {Griesshammer}}]{McGovern:2012ew}%
  \BibitemOpen
  \bibfield  {author} {\bibinfo {author} {\bibfnamefont {J.~A.}\ \bibnamefont
  {McGovern}}, \bibinfo {author} {\bibfnamefont {D.~R.}\ \bibnamefont
  {Phillips}}, \ and\ \bibinfo {author} {\bibfnamefont {H.~W.}\ \bibnamefont
  {Griesshammer}},\ }\bibfield  {title} {\enquote {\bibinfo {title} {{Compton
  scattering from the proton in an effective field theory with explicit Delta
  degrees of freedom}},}\ }\href {\doibase 10.1140/epja/i2013-13012-1}
  {\bibfield  {journal} {\bibinfo  {journal} {Eur. Phys. J. A}\ }\textbf
  {\bibinfo {volume} {49}},\ \bibinfo {pages} {12} (\bibinfo {year} {2013})},\
  \Eprint {http://arxiv.org/abs/1210.4104} {arXiv:1210.4104 [nucl-th]}
  \BibitemShut {NoStop}%
\bibitem [{\citenamefont {Griesshammer}\ \emph {et~al.}(2012)\citenamefont
  {Griesshammer}, \citenamefont {McGovern}, \citenamefont {Phillips},\ and\
  \citenamefont {Feldman}}]{Griesshammer:2012we}%
  \BibitemOpen
  \bibfield  {author} {\bibinfo {author} {\bibfnamefont {H.~W.}\ \bibnamefont
  {Griesshammer}}, \bibinfo {author} {\bibfnamefont {J.~A.}\ \bibnamefont
  {McGovern}}, \bibinfo {author} {\bibfnamefont {D.~R.}\ \bibnamefont
  {Phillips}}, \ and\ \bibinfo {author} {\bibfnamefont {G.}~\bibnamefont
  {Feldman}},\ }\bibfield  {title} {\enquote {\bibinfo {title} {{Using
  effective field theory to analyse low-energy Compton scattering data from
  protons and light nuclei}},}\ }\href {\doibase 10.1016/j.ppnp.2012.04.003}
  {\bibfield  {journal} {\bibinfo  {journal} {Prog. Part. Nucl. Phys.}\
  }\textbf {\bibinfo {volume} {67}},\ \bibinfo {pages} {841--897} (\bibinfo
  {year} {2012})},\ \Eprint {http://arxiv.org/abs/1203.6834} {arXiv:1203.6834
  [nucl-th]} \BibitemShut {NoStop}%
%%CITATION = ARXIV:1203.6834;%%
\bibitem [{\citenamefont {Fiebig}\ \emph {et~al.}(1989)\citenamefont {Fiebig},
  \citenamefont {Wilcox},\ and\ \citenamefont {Woloshyn}}]{Fiebig:1988en}%
  \BibitemOpen
  \bibfield  {author} {\bibinfo {author} {\bibfnamefont {H.~R.}\ \bibnamefont
  {Fiebig}}, \bibinfo {author} {\bibfnamefont {W.}~\bibnamefont {Wilcox}}, \
  and\ \bibinfo {author} {\bibfnamefont {R.~M.}\ \bibnamefont {Woloshyn}},\
  }\bibfield  {title} {\enquote {\bibinfo {title} {{A Study of Hadron Electric
  Polarizability in Quenched Lattice QCD}},}\ }\href {\doibase
  10.1016/0550-3213(89)90180-6} {\bibfield  {journal} {\bibinfo  {journal}
  {Nucl. Phys. B}\ }\textbf {\bibinfo {volume} {324}},\ \bibinfo {pages}
  {47--66} (\bibinfo {year} {1989})}\BibitemShut {NoStop}%
\bibitem [{\citenamefont {Lujan}\ \emph {et~al.}(2016)\citenamefont {Lujan},
  \citenamefont {Alexandru}, \citenamefont {Freeman},\ and\ \citenamefont
  {Lee}}]{Lujan:2016ffj}%
  \BibitemOpen
  \bibfield  {author} {\bibinfo {author} {\bibfnamefont {M.}~\bibnamefont
  {Lujan}}, \bibinfo {author} {\bibfnamefont {A.}~\bibnamefont {Alexandru}},
  \bibinfo {author} {\bibfnamefont {W.}~\bibnamefont {Freeman}}, \ and\
  \bibinfo {author} {\bibfnamefont {F.~X.}\ \bibnamefont {Lee}},\ }\bibfield
  {title} {\enquote {\bibinfo {title} {{Finite volume effects on the electric
  polarizability of neutral hadrons in lattice QCD}},}\ }\href {\doibase
  10.1103/PhysRevD.94.074506} {\bibfield  {journal} {\bibinfo  {journal} {Phys.
  Rev.}\ }\textbf {\bibinfo {volume} {D94}},\ \bibinfo {pages} {074506}
  (\bibinfo {year} {2016})},\ \Eprint {http://arxiv.org/abs/1606.07928}
  {arXiv:1606.07928 [hep-lat]} \BibitemShut {NoStop}%
%%CITATION = ARXIV:1606.07928;%%
\bibitem [{\citenamefont {Lujan}\ \emph {et~al.}(2014)\citenamefont {Lujan},
  \citenamefont {Alexandru}, \citenamefont {Freeman},\ and\ \citenamefont
  {Lee}}]{Lujan:2014kia}%
  \BibitemOpen
  \bibfield  {author} {\bibinfo {author} {\bibfnamefont {Michael}\ \bibnamefont
  {Lujan}}, \bibinfo {author} {\bibfnamefont {Andrei}\ \bibnamefont
  {Alexandru}}, \bibinfo {author} {\bibfnamefont {Walter}\ \bibnamefont
  {Freeman}}, \ and\ \bibinfo {author} {\bibfnamefont {Frank}\ \bibnamefont
  {Lee}},\ }\bibfield  {title} {\enquote {\bibinfo {title} {{Electric
  polarizability of neutral hadrons from dynamical lattice QCD ensembles}},}\
  }\href {\doibase 10.1103/PhysRevD.89.074506} {\bibfield  {journal} {\bibinfo
  {journal} {Phys. Rev.}\ }\textbf {\bibinfo {volume} {D89}},\ \bibinfo {pages}
  {074506} (\bibinfo {year} {2014})},\ \Eprint {http://arxiv.org/abs/1402.3025}
  {arXiv:1402.3025 [hep-lat]} \BibitemShut {NoStop}%
%%CITATION = ARXIV:1402.3025;%%
\bibitem [{\citenamefont {Freeman}\ \emph {et~al.}(2014)\citenamefont
  {Freeman}, \citenamefont {Alexandru}, \citenamefont {Lujan},\ and\
  \citenamefont {Lee}}]{Freeman:2014kka}%
  \BibitemOpen
  \bibfield  {author} {\bibinfo {author} {\bibfnamefont {Walter}\ \bibnamefont
  {Freeman}}, \bibinfo {author} {\bibfnamefont {Andrei}\ \bibnamefont
  {Alexandru}}, \bibinfo {author} {\bibfnamefont {Michael}\ \bibnamefont
  {Lujan}}, \ and\ \bibinfo {author} {\bibfnamefont {Frank~X.}\ \bibnamefont
  {Lee}},\ }\bibfield  {title} {\enquote {\bibinfo {title} {{Sea quark
  contributions to the electric polarizability of hadrons}},}\ }\href {\doibase
  10.1103/PhysRevD.90.054507} {\bibfield  {journal} {\bibinfo  {journal} {Phys.
  Rev. D}\ }\textbf {\bibinfo {volume} {90}},\ \bibinfo {pages} {054507}
  (\bibinfo {year} {2014})},\ \Eprint {http://arxiv.org/abs/1407.2687}
  {arXiv:1407.2687 [hep-lat]} \BibitemShut {NoStop}%
\bibitem [{\citenamefont {Freeman}\ \emph {et~al.}(2013)\citenamefont
  {Freeman}, \citenamefont {Alexandru}, \citenamefont {Lee},\ and\
  \citenamefont {Lujan}}]{Freeman:2013eta}%
  \BibitemOpen
  \bibfield  {author} {\bibinfo {author} {\bibfnamefont {Walter}\ \bibnamefont
  {Freeman}}, \bibinfo {author} {\bibfnamefont {Andrei}\ \bibnamefont
  {Alexandru}}, \bibinfo {author} {\bibfnamefont {Frank~X.}\ \bibnamefont
  {Lee}}, \ and\ \bibinfo {author} {\bibfnamefont {Mike}\ \bibnamefont
  {Lujan}},\ }\bibfield  {title} {\enquote {\bibinfo {title} {{Update on the
  Sea Contributions to Hadron Electric Polarizabilities through
  Reweighting}},}\ }in\ \href@noop {} {\emph {\bibinfo {booktitle} {{31st
  International Symposium on Lattice Field Theory}}}}\ (\bibinfo {year}
  {2013})\ \Eprint {http://arxiv.org/abs/1310.4426} {arXiv:1310.4426 [hep-lat]}
  \BibitemShut {NoStop}%
\bibitem [{\citenamefont {Tiburzi}(2008)}]{Tiburzi:2008ma}%
  \BibitemOpen
  \bibfield  {author} {\bibinfo {author} {\bibfnamefont {Brian~C.}\
  \bibnamefont {Tiburzi}},\ }\bibfield  {title} {\enquote {\bibinfo {title}
  {{Hadrons in Strong Electric and Magnetic Fields}},}\ }\href {\doibase
  10.1016/j.nuclphysa.2008.10.010} {\bibfield  {journal} {\bibinfo  {journal}
  {Nucl. Phys.}\ }\textbf {\bibinfo {volume} {A814}},\ \bibinfo {pages}
  {74--108} (\bibinfo {year} {2008})},\ \Eprint
  {http://arxiv.org/abs/0808.3965} {arXiv:0808.3965 [hep-ph]} \BibitemShut
  {NoStop}%
%%CITATION = ARXIV:0808.3965;%%
\bibitem [{\citenamefont {Detmold}\ \emph {et~al.}(2009)\citenamefont
  {Detmold}, \citenamefont {Tiburzi},\ and\ \citenamefont
  {Walker-Loud}}]{Detmold:2009fr}%
  \BibitemOpen
  \bibfield  {author} {\bibinfo {author} {\bibfnamefont {William}\ \bibnamefont
  {Detmold}}, \bibinfo {author} {\bibfnamefont {Brian~C.}\ \bibnamefont
  {Tiburzi}}, \ and\ \bibinfo {author} {\bibfnamefont {Andre}\ \bibnamefont
  {Walker-Loud}},\ }\bibfield  {title} {\enquote {\bibinfo {title} {{Lattice
  QCD in Background Fields}},}\ }\bibfield  {booktitle} {\emph {\bibinfo
  {booktitle} {{Proceedings, 10th Workshop on Non-Perturbative Quantum
  Chromodynamics : Paris, France, June 8-12, 2009}}},\ }\href@noop {} {\
  (\bibinfo {year} {2009})},\ \Eprint {http://arxiv.org/abs/0908.3626}
  {arXiv:0908.3626 [hep-lat]} \BibitemShut {NoStop}%
%%CITATION = ARXIV:0908.3626;%%
\bibitem [{\citenamefont {Alexandru}\ and\ \citenamefont
  {Lee}(2008)}]{Alexandru:2008sj}%
  \BibitemOpen
  \bibfield  {author} {\bibinfo {author} {\bibfnamefont {Andrei}\ \bibnamefont
  {Alexandru}}\ and\ \bibinfo {author} {\bibfnamefont {Frank~X.}\ \bibnamefont
  {Lee}},\ }\bibfield  {title} {\enquote {\bibinfo {title} {{The Background
  field method on the lattice}},}\ }\href {\doibase 10.22323/1.066.0145}
  {\bibfield  {journal} {\bibinfo  {journal} {PoS}\ }\textbf {\bibinfo {volume}
  {LATTICE2008}},\ \bibinfo {pages} {145} (\bibinfo {year} {2008})},\ \Eprint
  {http://arxiv.org/abs/0810.2833} {arXiv:0810.2833 [hep-lat]} \BibitemShut
  {NoStop}%
\bibitem [{\citenamefont {Lee}\ \emph {et~al.}(2006)\citenamefont {Lee},
  \citenamefont {Zhou}, \citenamefont {Wilcox},\ and\ \citenamefont
  {Christensen}}]{Lee:2005dq}%
  \BibitemOpen
  \bibfield  {author} {\bibinfo {author} {\bibfnamefont {Frank~X.}\
  \bibnamefont {Lee}}, \bibinfo {author} {\bibfnamefont {Leming}\ \bibnamefont
  {Zhou}}, \bibinfo {author} {\bibfnamefont {Walter}\ \bibnamefont {Wilcox}}, \
  and\ \bibinfo {author} {\bibfnamefont {Joseph~C.}\ \bibnamefont
  {Christensen}},\ }\bibfield  {title} {\enquote {\bibinfo {title} {{Magnetic
  polarizability of hadrons from lattice QCD in the background field
  method}},}\ }\href {\doibase 10.1103/PhysRevD.73.034503} {\bibfield
  {journal} {\bibinfo  {journal} {Phys. Rev. D}\ }\textbf {\bibinfo {volume}
  {73}},\ \bibinfo {pages} {034503} (\bibinfo {year} {2006})},\ \Eprint
  {http://arxiv.org/abs/hep-lat/0509065} {arXiv:hep-lat/0509065} \BibitemShut
  {NoStop}%
\bibitem [{\citenamefont {Lee}\ \emph {et~al.}(2005)\citenamefont {Lee},
  \citenamefont {Kelly}, \citenamefont {Zhou},\ and\ \citenamefont
  {Wilcox}}]{Lee:2005ds}%
  \BibitemOpen
  \bibfield  {author} {\bibinfo {author} {\bibfnamefont {F.~X.}\ \bibnamefont
  {Lee}}, \bibinfo {author} {\bibfnamefont {R.}~\bibnamefont {Kelly}}, \bibinfo
  {author} {\bibfnamefont {L.}~\bibnamefont {Zhou}}, \ and\ \bibinfo {author}
  {\bibfnamefont {W.}~\bibnamefont {Wilcox}},\ }\bibfield  {title} {\enquote
  {\bibinfo {title} {{Baryon magnetic moments in the background field
  method}},}\ }\href {\doibase 10.1016/j.physletb.2005.08.106} {\bibfield
  {journal} {\bibinfo  {journal} {Phys. Lett. B}\ }\textbf {\bibinfo {volume}
  {627}},\ \bibinfo {pages} {71--76} (\bibinfo {year} {2005})},\ \Eprint
  {http://arxiv.org/abs/hep-lat/0509067} {arXiv:hep-lat/0509067} \BibitemShut
  {NoStop}%
\bibitem [{\citenamefont {Engelhardt}(2007)}]{Engelhardt:2007ub}%
  \BibitemOpen
  \bibfield  {author} {\bibinfo {author} {\bibfnamefont {Michael}\ \bibnamefont
  {Engelhardt}},\ }\bibfield  {title} {\enquote {\bibinfo {title} {{Neutron
  electric polarizability from unquenched lattice QCD using the background
  field approach}},}\ }\href {\doibase 10.1103/PhysRevD.76.114502} {\bibfield
  {journal} {\bibinfo  {journal} {Phys. Rev. D}\ }\textbf {\bibinfo {volume}
  {76}},\ \bibinfo {pages} {114502} (\bibinfo {year} {2007})},\ \Eprint
  {http://arxiv.org/abs/0706.3919} {arXiv:0706.3919 [hep-lat]} \BibitemShut
  {NoStop}%
\bibitem [{\citenamefont {Bignell}\ \emph
  {et~al.}(2020{\natexlab{a}})\citenamefont {Bignell}, \citenamefont {Kamleh},\
  and\ \citenamefont {Leinweber}}]{Bignell:2020xkf}%
  \BibitemOpen
  \bibfield  {author} {\bibinfo {author} {\bibfnamefont {Ryan}\ \bibnamefont
  {Bignell}}, \bibinfo {author} {\bibfnamefont {Waseem}\ \bibnamefont
  {Kamleh}}, \ and\ \bibinfo {author} {\bibfnamefont {Derek}\ \bibnamefont
  {Leinweber}},\ }\bibfield  {title} {\enquote {\bibinfo {title} {{Magnetic
  polarizability of the nucleon using a Laplacian mode projection}},}\ }\href
  {\doibase 10.1103/PhysRevD.101.094502} {\bibfield  {journal} {\bibinfo
  {journal} {Phys. Rev. D}\ }\textbf {\bibinfo {volume} {101}},\ \bibinfo
  {pages} {094502} (\bibinfo {year} {2020}{\natexlab{a}})},\ \Eprint
  {http://arxiv.org/abs/2002.07915} {arXiv:2002.07915 [hep-lat]} \BibitemShut
  {NoStop}%
\bibitem [{\citenamefont {Deshmukh}\ and\ \citenamefont
  {Tiburzi}(2018)}]{Deshmukh:2017ciw}%
  \BibitemOpen
  \bibfield  {author} {\bibinfo {author} {\bibfnamefont {Amol}\ \bibnamefont
  {Deshmukh}}\ and\ \bibinfo {author} {\bibfnamefont {Brian~C.}\ \bibnamefont
  {Tiburzi}},\ }\bibfield  {title} {\enquote {\bibinfo {title} {{Octet Baryons
  in Large Magnetic Fields}},}\ }\href {\doibase 10.1103/PhysRevD.97.014006}
  {\bibfield  {journal} {\bibinfo  {journal} {Phys. Rev. D}\ }\textbf {\bibinfo
  {volume} {97}},\ \bibinfo {pages} {014006} (\bibinfo {year} {2018})},\
  \Eprint {http://arxiv.org/abs/1709.04997} {arXiv:1709.04997 [hep-ph]}
  \BibitemShut {NoStop}%
\bibitem [{\citenamefont {Bali}\ \emph
  {et~al.}(2018{\natexlab{a}})\citenamefont {Bali}, \citenamefont {Brandt},
  \citenamefont {Endr\H{o}di},\ and\ \citenamefont
  {Gl\"a\ss{}le}}]{Bali:2017ian}%
  \BibitemOpen
  \bibfield  {author} {\bibinfo {author} {\bibfnamefont {Gunnar~S.}\
  \bibnamefont {Bali}}, \bibinfo {author} {\bibfnamefont {Bastian~B.}\
  \bibnamefont {Brandt}}, \bibinfo {author} {\bibfnamefont {Gergely}\
  \bibnamefont {Endr\H{o}di}}, \ and\ \bibinfo {author} {\bibfnamefont
  {Benjamin}\ \bibnamefont {Gl\"a\ss{}le}},\ }\bibfield  {title} {\enquote
  {\bibinfo {title} {{Meson masses in electromagnetic fields with Wilson
  fermions}},}\ }\href {\doibase 10.1103/PhysRevD.97.034505} {\bibfield
  {journal} {\bibinfo  {journal} {Phys. Rev. D}\ }\textbf {\bibinfo {volume}
  {97}},\ \bibinfo {pages} {034505} (\bibinfo {year} {2018}{\natexlab{a}})},\
  \Eprint {http://arxiv.org/abs/1707.05600} {arXiv:1707.05600 [hep-lat]}
  \BibitemShut {NoStop}%
\bibitem [{\citenamefont {Bruckmann}\ \emph {et~al.}(2017)\citenamefont
  {Bruckmann}, \citenamefont {Endrodi}, \citenamefont {Giordano}, \citenamefont
  {Katz}, \citenamefont {Kovacs}, \citenamefont {Pittler},\ and\ \citenamefont
  {Wellnhofer}}]{Bruckmann:2017pft}%
  \BibitemOpen
  \bibfield  {author} {\bibinfo {author} {\bibfnamefont {F.}~\bibnamefont
  {Bruckmann}}, \bibinfo {author} {\bibfnamefont {G.}~\bibnamefont {Endrodi}},
  \bibinfo {author} {\bibfnamefont {M.}~\bibnamefont {Giordano}}, \bibinfo
  {author} {\bibfnamefont {S.~D.}\ \bibnamefont {Katz}}, \bibinfo {author}
  {\bibfnamefont {T.~G.}\ \bibnamefont {Kovacs}}, \bibinfo {author}
  {\bibfnamefont {F.}~\bibnamefont {Pittler}}, \ and\ \bibinfo {author}
  {\bibfnamefont {J.}~\bibnamefont {Wellnhofer}},\ }\bibfield  {title}
  {\enquote {\bibinfo {title} {{Landau levels in QCD}},}\ }\href {\doibase
  10.1103/PhysRevD.96.074506} {\bibfield  {journal} {\bibinfo  {journal} {Phys.
  Rev. D}\ }\textbf {\bibinfo {volume} {96}},\ \bibinfo {pages} {074506}
  (\bibinfo {year} {2017})},\ \Eprint {http://arxiv.org/abs/1705.10210}
  {arXiv:1705.10210 [hep-lat]} \BibitemShut {NoStop}%
\bibitem [{\citenamefont {Parreno}\ \emph {et~al.}(2017)\citenamefont
  {Parreno}, \citenamefont {Savage}, \citenamefont {Tiburzi}, \citenamefont
  {Wilhelm}, \citenamefont {Chang}, \citenamefont {Detmold},\ and\
  \citenamefont {Orginos}}]{Parreno:2016fwu}%
  \BibitemOpen
  \bibfield  {author} {\bibinfo {author} {\bibfnamefont {Assumpta}\
  \bibnamefont {Parreno}}, \bibinfo {author} {\bibfnamefont {Martin~J.}\
  \bibnamefont {Savage}}, \bibinfo {author} {\bibfnamefont {Brian~C.}\
  \bibnamefont {Tiburzi}}, \bibinfo {author} {\bibfnamefont {Jonas}\
  \bibnamefont {Wilhelm}}, \bibinfo {author} {\bibfnamefont {Emmanuel}\
  \bibnamefont {Chang}}, \bibinfo {author} {\bibfnamefont {William}\
  \bibnamefont {Detmold}}, \ and\ \bibinfo {author} {\bibfnamefont {Kostas}\
  \bibnamefont {Orginos}},\ }\bibfield  {title} {\enquote {\bibinfo {title}
  {{Octet baryon magnetic moments from lattice QCD: Approaching experiment from
  a three-flavor symmetric point}},}\ }\href {\doibase
  10.1103/PhysRevD.95.114513} {\bibfield  {journal} {\bibinfo  {journal} {Phys.
  Rev. D}\ }\textbf {\bibinfo {volume} {95}},\ \bibinfo {pages} {114513}
  (\bibinfo {year} {2017})},\ \Eprint {http://arxiv.org/abs/1609.03985}
  {arXiv:1609.03985 [hep-lat]} \BibitemShut {NoStop}%
\bibitem [{\citenamefont {Luschevskaya}\ \emph {et~al.}(2016)\citenamefont
  {Luschevskaya}, \citenamefont {Solovjeva},\ and\ \citenamefont
  {Teryaev}}]{Luschevskaya:2015cko}%
  \BibitemOpen
  \bibfield  {author} {\bibinfo {author} {\bibfnamefont {E.~V.}\ \bibnamefont
  {Luschevskaya}}, \bibinfo {author} {\bibfnamefont {O.~E.}\ \bibnamefont
  {Solovjeva}}, \ and\ \bibinfo {author} {\bibfnamefont {O.~V.}\ \bibnamefont
  {Teryaev}},\ }\bibfield  {title} {\enquote {\bibinfo {title} {{Magnetic
  polarizability of pion}},}\ }\href {\doibase 10.1016/j.physletb.2016.08.054}
  {\bibfield  {journal} {\bibinfo  {journal} {Phys. Lett. B}\ }\textbf
  {\bibinfo {volume} {761}},\ \bibinfo {pages} {393--398} (\bibinfo {year}
  {2016})},\ \Eprint {http://arxiv.org/abs/1511.09316} {arXiv:1511.09316
  [hep-lat]} \BibitemShut {NoStop}%
\bibitem [{\citenamefont {Chang}\ \emph {et~al.}(2015)\citenamefont {Chang},
  \citenamefont {Detmold}, \citenamefont {Orginos}, \citenamefont {Parreno},
  \citenamefont {Savage}, \citenamefont {Tiburzi},\ and\ \citenamefont
  {Beane}}]{Chang:2015qxa}%
  \BibitemOpen
  \bibfield  {author} {\bibinfo {author} {\bibfnamefont {Emmanuel}\
  \bibnamefont {Chang}}, \bibinfo {author} {\bibfnamefont {William}\
  \bibnamefont {Detmold}}, \bibinfo {author} {\bibfnamefont {Kostas}\
  \bibnamefont {Orginos}}, \bibinfo {author} {\bibfnamefont {Assumpta}\
  \bibnamefont {Parreno}}, \bibinfo {author} {\bibfnamefont {Martin~J.}\
  \bibnamefont {Savage}}, \bibinfo {author} {\bibfnamefont {Brian~C.}\
  \bibnamefont {Tiburzi}}, \ and\ \bibinfo {author} {\bibfnamefont {Silas~R.}\
  \bibnamefont {Beane}} (\bibinfo {collaboration} {NPLQCD}),\ }\bibfield
  {title} {\enquote {\bibinfo {title} {{Magnetic structure of light nuclei from
  lattice QCD}},}\ }\href {\doibase 10.1103/PhysRevD.92.114502} {\bibfield
  {journal} {\bibinfo  {journal} {Phys. Rev. D}\ }\textbf {\bibinfo {volume}
  {92}},\ \bibinfo {pages} {114502} (\bibinfo {year} {2015})},\ \Eprint
  {http://arxiv.org/abs/1506.05518} {arXiv:1506.05518 [hep-lat]} \BibitemShut
  {NoStop}%
\bibitem [{\citenamefont {Detmold}\ \emph {et~al.}(2010)\citenamefont
  {Detmold}, \citenamefont {Tiburzi},\ and\ \citenamefont
  {Walker-Loud}}]{Detmold:2010ts}%
  \BibitemOpen
  \bibfield  {author} {\bibinfo {author} {\bibfnamefont {W.}~\bibnamefont
  {Detmold}}, \bibinfo {author} {\bibfnamefont {B.~C.}\ \bibnamefont
  {Tiburzi}}, \ and\ \bibinfo {author} {\bibfnamefont {A.}~\bibnamefont
  {Walker-Loud}},\ }\bibfield  {title} {\enquote {\bibinfo {title} {{Extracting
  Nucleon Magnetic Moments and Electric Polarizabilities from Lattice QCD in
  Background Electric Fields}},}\ }\href {\doibase 10.1103/PhysRevD.81.054502}
  {\bibfield  {journal} {\bibinfo  {journal} {Phys. Rev.}\ }\textbf {\bibinfo
  {volume} {D81}},\ \bibinfo {pages} {054502} (\bibinfo {year} {2010})},\
  \Eprint {http://arxiv.org/abs/1001.1131} {arXiv:1001.1131 [hep-lat]}
  \BibitemShut {NoStop}%
%%CITATION = ARXIV:1001.1131;%%
\bibitem [{\citenamefont {Davoudi}\ and\ \citenamefont
  {Detmold}(2015)}]{Davoudi:2015cba}%
  \BibitemOpen
  \bibfield  {author} {\bibinfo {author} {\bibfnamefont {Zohreh}\ \bibnamefont
  {Davoudi}}\ and\ \bibinfo {author} {\bibfnamefont {William}\ \bibnamefont
  {Detmold}},\ }\bibfield  {title} {\enquote {\bibinfo {title} {{Implementation
  of general background electromagnetic fields on a periodic hypercubic
  lattice}},}\ }\href {\doibase 10.1103/PhysRevD.92.074506} {\bibfield
  {journal} {\bibinfo  {journal} {Phys. Rev. D}\ }\textbf {\bibinfo {volume}
  {92}},\ \bibinfo {pages} {074506} (\bibinfo {year} {2015})},\ \Eprint
  {http://arxiv.org/abs/1507.01908} {arXiv:1507.01908 [hep-lat]} \BibitemShut
  {NoStop}%
\bibitem [{\citenamefont {Engelhardt}(2011)}]{Engelhardt:2011qq}%
  \BibitemOpen
  \bibfield  {author} {\bibinfo {author} {\bibfnamefont {Michael}\ \bibnamefont
  {Engelhardt}},\ }\bibfield  {title} {\enquote {\bibinfo {title} {{Exploration
  of the electric spin polarizability of the neutron in lattice QCD}},}\ }\href
  {\doibase 10.22323/1.139.0153} {\bibfield  {journal} {\bibinfo  {journal}
  {PoS}\ }\textbf {\bibinfo {volume} {LATTICE2011}},\ \bibinfo {pages} {153}
  (\bibinfo {year} {2011})},\ \Eprint {http://arxiv.org/abs/1111.3686}
  {arXiv:1111.3686 [hep-lat]} \BibitemShut {NoStop}%
\bibitem [{\citenamefont {Lee}\ and\ \citenamefont
  {Alexandru}(2011)}]{Lee:2011gz}%
  \BibitemOpen
  \bibfield  {author} {\bibinfo {author} {\bibfnamefont {Frank~X.}\
  \bibnamefont {Lee}}\ and\ \bibinfo {author} {\bibfnamefont {Andrei}\
  \bibnamefont {Alexandru}},\ }\bibfield  {title} {\enquote {\bibinfo {title}
  {{Spin Polarizabilities on the Lattice}},}\ }\bibfield  {booktitle} {\emph
  {\bibinfo {booktitle} {{Proceedings, 29th International Symposium on Lattice
  field theory (Lattice 2011): Squaw Valley, Lake Tahoe, USA, July 10-16,
  2011}}},\ }\href {\doibase 10.22323/1.139.0317} {\bibfield  {journal}
  {\bibinfo  {journal} {PoS}\ }\textbf {\bibinfo {volume} {LATTICE2011}},\
  \bibinfo {pages} {317} (\bibinfo {year} {2011})},\ \Eprint
  {http://arxiv.org/abs/1111.4425} {arXiv:1111.4425 [hep-lat]} \BibitemShut
  {NoStop}%
%%CITATION = ARXIV:1111.4425;%%
\bibitem [{\citenamefont {Detmold}\ \emph {et~al.}(2006)\citenamefont
  {Detmold}, \citenamefont {Tiburzi},\ and\ \citenamefont
  {Walker-Loud}}]{Detmold:2006vu}%
  \BibitemOpen
  \bibfield  {author} {\bibinfo {author} {\bibfnamefont {W.}~\bibnamefont
  {Detmold}}, \bibinfo {author} {\bibfnamefont {B.~C.}\ \bibnamefont
  {Tiburzi}}, \ and\ \bibinfo {author} {\bibfnamefont {Andre}\ \bibnamefont
  {Walker-Loud}},\ }\bibfield  {title} {\enquote {\bibinfo {title}
  {{Electromagnetic and spin polarisabilities in lattice QCD}},}\ }\href
  {\doibase 10.1103/PhysRevD.73.114505} {\bibfield  {journal} {\bibinfo
  {journal} {Phys. Rev.}\ }\textbf {\bibinfo {volume} {D73}},\ \bibinfo {pages}
  {114505} (\bibinfo {year} {2006})},\ \Eprint
  {http://arxiv.org/abs/hep-lat/0603026} {arXiv:hep-lat/0603026 [hep-lat]}
  \BibitemShut {NoStop}%
%%CITATION = HEP-LAT/0603026;%%
\bibitem [{\citenamefont {Niyazi}\ \emph {et~al.}(2021)\citenamefont {Niyazi},
  \citenamefont {Alexandru}, \citenamefont {Lee},\ and\ \citenamefont
  {Lujan}}]{niyazi2021charged}%
  \BibitemOpen
  \bibfield  {author} {\bibinfo {author} {\bibfnamefont {Hossein}\ \bibnamefont
  {Niyazi}}, \bibinfo {author} {\bibfnamefont {Andrei}\ \bibnamefont
  {Alexandru}}, \bibinfo {author} {\bibfnamefont {Frank~X.}\ \bibnamefont
  {Lee}}, \ and\ \bibinfo {author} {\bibfnamefont {Michael}\ \bibnamefont
  {Lujan}},\ }\href@noop {} {\enquote {\bibinfo {title} {Charged pion electric
  polarizability from lattice qcd},}\ } (\bibinfo {year} {2021}),\ \Eprint
  {http://arxiv.org/abs/2105.06906} {arXiv:2105.06906 [hep-lat]} \BibitemShut
  {NoStop}%
\bibitem [{\citenamefont {Bignell}\ \emph
  {et~al.}(2020{\natexlab{b}})\citenamefont {Bignell}, \citenamefont {Kamleh},\
  and\ \citenamefont {Leinweber}}]{Bignell_2020}%
  \BibitemOpen
  \bibfield  {author} {\bibinfo {author} {\bibfnamefont {Ryan}\ \bibnamefont
  {Bignell}}, \bibinfo {author} {\bibfnamefont {Waseem}\ \bibnamefont
  {Kamleh}}, \ and\ \bibinfo {author} {\bibfnamefont {Derek}\ \bibnamefont
  {Leinweber}},\ }\bibfield  {title} {\enquote {\bibinfo {title} {Pion magnetic
  polarisability using the background field method},}\ }\href {\doibase
  10.1016/j.physletb.2020.135853} {\bibfield  {journal} {\bibinfo  {journal}
  {Physics Letters B}\ }\textbf {\bibinfo {volume} {811}},\ \bibinfo {pages}
  {135853} (\bibinfo {year} {2020}{\natexlab{b}})}\BibitemShut {NoStop}%
\bibitem [{\citenamefont {He}\ \emph {et~al.}(2021)\citenamefont {He},
  \citenamefont {Leinweber}, \citenamefont {Thomas},\ and\ \citenamefont
  {Wang}}]{He:2021eha}%
  \BibitemOpen
  \bibfield  {author} {\bibinfo {author} {\bibfnamefont {Fangcheng}\
  \bibnamefont {He}}, \bibinfo {author} {\bibfnamefont {Derek~B.}\ \bibnamefont
  {Leinweber}}, \bibinfo {author} {\bibfnamefont {Anthony~W.}\ \bibnamefont
  {Thomas}}, \ and\ \bibinfo {author} {\bibfnamefont {Ping}\ \bibnamefont
  {Wang}},\ }\bibfield  {title} {\enquote {\bibinfo {title} {{Chiral
  extrapolation of the charged-pion magnetic polarizability with Pad\'e
  approximant}},}\ }\href@noop {} {\  (\bibinfo {year} {2021})},\ \Eprint
  {http://arxiv.org/abs/2104.09963} {arXiv:2104.09963 [nucl-th]} \BibitemShut
  {NoStop}%
\bibitem [{\citenamefont {Liang}\ \emph {et~al.}(2020)\citenamefont {Liang},
  \citenamefont {Draper}, \citenamefont {Liu}, \citenamefont {Rothkopf},\ and\
  \citenamefont {Yang}}]{Liang:2019frk}%
  \BibitemOpen
  \bibfield  {author} {\bibinfo {author} {\bibfnamefont {Jian}\ \bibnamefont
  {Liang}}, \bibinfo {author} {\bibfnamefont {Terrence}\ \bibnamefont
  {Draper}}, \bibinfo {author} {\bibfnamefont {Keh-Fei}\ \bibnamefont {Liu}},
  \bibinfo {author} {\bibfnamefont {Alexander}\ \bibnamefont {Rothkopf}}, \
  and\ \bibinfo {author} {\bibfnamefont {Yi-Bo}\ \bibnamefont {Yang}} (\bibinfo
  {collaboration} {XQCD}),\ }\bibfield  {title} {\enquote {\bibinfo {title}
  {{Towards the nucleon hadronic tensor from lattice QCD}},}\ }\href {\doibase
  10.1103/PhysRevD.101.114503} {\bibfield  {journal} {\bibinfo  {journal}
  {Phys. Rev. D}\ }\textbf {\bibinfo {volume} {101}},\ \bibinfo {pages}
  {114503} (\bibinfo {year} {2020})},\ \Eprint
  {http://arxiv.org/abs/1906.05312} {arXiv:1906.05312 [hep-ph]} \BibitemShut
  {NoStop}%
\bibitem [{\citenamefont {Liang}\ and\ \citenamefont
  {Liu}(2020)}]{Liang_2020a}%
  \BibitemOpen
  \bibfield  {author} {\bibinfo {author} {\bibfnamefont {Jian}\ \bibnamefont
  {Liang}}\ and\ \bibinfo {author} {\bibfnamefont {Keh-Fei}\ \bibnamefont
  {Liu}},\ }\bibfield  {title} {\enquote {\bibinfo {title} {Pdfs and
  neutrino-nucleon scattering from hadronic tensor},}\ }\href {\doibase
  10.22323/1.363.0046} {\bibfield  {journal} {\bibinfo  {journal} {Proceedings
  of 37th International Symposium on Lattice Field Theory ---
  PoS(LATTICE2019)}\ } (\bibinfo {year} {2020}),\
  10.22323/1.363.0046}\BibitemShut {NoStop}%
\bibitem [{\citenamefont {Fu}(2012)}]{Fu_2012}%
  \BibitemOpen
  \bibfield  {author} {\bibinfo {author} {\bibfnamefont {Ziwen}\ \bibnamefont
  {Fu}},\ }\bibfield  {title} {\enquote {\bibinfo {title} {Lattice study on
  $\ensuremath{\pi}k$ scattering with moving wall source},}\ }\href {\doibase
  10.1103/PhysRevD.85.074501} {\bibfield  {journal} {\bibinfo  {journal} {Phys.
  Rev. D}\ }\textbf {\bibinfo {volume} {85}},\ \bibinfo {pages} {074501}
  (\bibinfo {year} {2012})}\BibitemShut {NoStop}%
\bibitem [{\citenamefont {Alexandrou}(2004)}]{Alexandrou_2004}%
  \BibitemOpen
  \bibfield  {author} {\bibinfo {author} {\bibfnamefont {C.}~\bibnamefont
  {Alexandrou}},\ }\bibfield  {title} {\enquote {\bibinfo {title} {Hadron
  deformation from lattice qcd},}\ }\href {\doibase
  10.1016/s0920-5632(03)02451-4} {\bibfield  {journal} {\bibinfo  {journal}
  {Nuclear Physics B - Proceedings Supplements}\ }\textbf {\bibinfo {volume}
  {128}},\ \bibinfo {pages} {1--8} (\bibinfo {year} {2004})}\BibitemShut
  {NoStop}%
\bibitem [{\citenamefont {Bali}\ \emph
  {et~al.}(2018{\natexlab{b}})\citenamefont {Bali}, \citenamefont {Bruns},
  \citenamefont {Castagnini}, \citenamefont {Diehl}, \citenamefont {Gaunt},
  \citenamefont {Gl{\"a}{\ss}le}, \citenamefont {Sch{\"a}fer}, \citenamefont
  {Sternbeck},\ and\ \citenamefont {Zimmermann}}]{Bali_2018}%
  \BibitemOpen
  \bibfield  {author} {\bibinfo {author} {\bibfnamefont {Gunnar~S.}\
  \bibnamefont {Bali}}, \bibinfo {author} {\bibfnamefont {Peter~C.}\
  \bibnamefont {Bruns}}, \bibinfo {author} {\bibfnamefont {Luca}\ \bibnamefont
  {Castagnini}}, \bibinfo {author} {\bibfnamefont {Markus}\ \bibnamefont
  {Diehl}}, \bibinfo {author} {\bibfnamefont {Jonathan~R.}\ \bibnamefont
  {Gaunt}}, \bibinfo {author} {\bibfnamefont {Benjamin}\ \bibnamefont
  {Gl{\"a}{\ss}le}}, \bibinfo {author} {\bibfnamefont {Andreas}\ \bibnamefont
  {Sch{\"a}fer}}, \bibinfo {author} {\bibfnamefont {Andr{\'e}}\ \bibnamefont
  {Sternbeck}}, \ and\ \bibinfo {author} {\bibfnamefont {Christian}\
  \bibnamefont {Zimmermann}},\ }\bibfield  {title} {\enquote {\bibinfo {title}
  {Two-current correlations in the pion on the lattice},}\ }\href {\doibase
  10.1007/jhep12(2018)061} {\bibfield  {journal} {\bibinfo  {journal} {Journal
  of High Energy Physics}\ }\textbf {\bibinfo {volume} {2018}} (\bibinfo {year}
  {2018}{\natexlab{b}}),\ 10.1007/jhep12(2018)061}\BibitemShut {NoStop}%
\bibitem [{\citenamefont {Bali}\ \emph {et~al.}(2021)\citenamefont {Bali},
  \citenamefont {Diehl}, \citenamefont {Gl{\"a}{\ss}le}, \citenamefont
  {Sch{\"a}fer},\ and\ \citenamefont {Zimmermann}}]{bali2021double}%
  \BibitemOpen
  \bibfield  {author} {\bibinfo {author} {\bibfnamefont {Gunnar~S.}\
  \bibnamefont {Bali}}, \bibinfo {author} {\bibfnamefont {Markus}\ \bibnamefont
  {Diehl}}, \bibinfo {author} {\bibfnamefont {Benjamin}\ \bibnamefont
  {Gl{\"a}{\ss}le}}, \bibinfo {author} {\bibfnamefont {Andreas}\ \bibnamefont
  {Sch{\"a}fer}}, \ and\ \bibinfo {author} {\bibfnamefont {Christian}\
  \bibnamefont {Zimmermann}},\ }\href@noop {} {\enquote {\bibinfo {title}
  {Double parton distributions in the nucleon from lattice qcd},}\ } (\bibinfo
  {year} {2021}),\ \Eprint {http://arxiv.org/abs/2106.03451} {arXiv:2106.03451
  [hep-lat]} \BibitemShut {NoStop}%
\bibitem [{\citenamefont {Burkardt}\ \emph {et~al.}(1995)\citenamefont
  {Burkardt}, \citenamefont {Grandy},\ and\ \citenamefont
  {Negele}}]{BURKARDT1995441}%
  \BibitemOpen
  \bibfield  {author} {\bibinfo {author} {\bibfnamefont {M.}~\bibnamefont
  {Burkardt}}, \bibinfo {author} {\bibfnamefont {J.M.}\ \bibnamefont {Grandy}},
  \ and\ \bibinfo {author} {\bibfnamefont {J.W.}\ \bibnamefont {Negele}},\
  }\bibfield  {title} {\enquote {\bibinfo {title} {Calculation and
  interpretation of hadron correlation functions in lattice qcd},}\ }\href
  {\doibase https://doi.org/10.1006/aphy.1995.1026} {\bibfield  {journal}
  {\bibinfo  {journal} {Annals of Physics}\ }\textbf {\bibinfo {volume}
  {238}},\ \bibinfo {pages} {441--472} (\bibinfo {year} {1995})}\BibitemShut
  {NoStop}%
\bibitem [{\citenamefont {Wilcox}(1997)}]{Wilcox:1996vx}%
  \BibitemOpen
  \bibfield  {author} {\bibinfo {author} {\bibfnamefont {Walter}\ \bibnamefont
  {Wilcox}},\ }\bibfield  {title} {\enquote {\bibinfo {title} {{Lattice charge
  overlap. 2: Aspects of charged pion polarizability}},}\ }\href {\doibase
  10.1006/aphy.1996.5649} {\bibfield  {journal} {\bibinfo  {journal} {Annals
  Phys.}\ }\textbf {\bibinfo {volume} {255}},\ \bibinfo {pages} {60--74}
  (\bibinfo {year} {1997})},\ \Eprint {http://arxiv.org/abs/hep-lat/9606019}
  {arXiv:hep-lat/9606019} \BibitemShut {NoStop}%
\bibitem [{\citenamefont {Gasser}\ \emph {et~al.}(2020)\citenamefont {Gasser},
  \citenamefont {Hoferichter}, \citenamefont {Leutwyler},\ and\ \citenamefont
  {Rusetsky}}]{gasser2020cottingham}%
  \BibitemOpen
  \bibfield  {author} {\bibinfo {author} {\bibfnamefont {J.}~\bibnamefont
  {Gasser}}, \bibinfo {author} {\bibfnamefont {M.}~\bibnamefont {Hoferichter}},
  \bibinfo {author} {\bibfnamefont {H.}~\bibnamefont {Leutwyler}}, \ and\
  \bibinfo {author} {\bibfnamefont {A.}~\bibnamefont {Rusetsky}},\ }\href@noop
  {} {\enquote {\bibinfo {title} {Cottingham formula and nucleon
  polarizabilities},}\ } (\bibinfo {year} {2020}),\ \Eprint
  {http://arxiv.org/abs/1506.06747} {arXiv:1506.06747 [hep-ph]} \BibitemShut
  {NoStop}%
\bibitem [{\citenamefont {Kuramashi}\ \emph {et~al.}(1993)\citenamefont
  {Kuramashi}, \citenamefont {Fukugita}, \citenamefont {Mino}, \citenamefont
  {Okawa},\ and\ \citenamefont {Ukawa}}]{WallSource1993}%
  \BibitemOpen
  \bibfield  {author} {\bibinfo {author} {\bibfnamefont {Y.}~\bibnamefont
  {Kuramashi}}, \bibinfo {author} {\bibfnamefont {M.}~\bibnamefont {Fukugita}},
  \bibinfo {author} {\bibfnamefont {H.}~\bibnamefont {Mino}}, \bibinfo {author}
  {\bibfnamefont {M.}~\bibnamefont {Okawa}}, \ and\ \bibinfo {author}
  {\bibfnamefont {A.}~\bibnamefont {Ukawa}},\ }\bibfield  {title} {\enquote
  {\bibinfo {title} {Lattice qcd calculation of full pion scattering
  lengths},}\ }\href {\doibase 10.1103/PhysRevLett.71.2387} {\bibfield
  {journal} {\bibinfo  {journal} {Phys. Rev. Lett.}\ }\textbf {\bibinfo
  {volume} {71}},\ \bibinfo {pages} {2387--2390} (\bibinfo {year}
  {1993})}\BibitemShut {NoStop}%
\bibitem [{\citenamefont {Wilcox}(1992)}]{Wilcox_1992}%
  \BibitemOpen
  \bibfield  {author} {\bibinfo {author} {\bibfnamefont {Walter}\ \bibnamefont
  {Wilcox}},\ }\bibfield  {title} {\enquote {\bibinfo {title} {Lattice charge
  overlap: towards the elastic limit},}\ }\href {\doibase
  10.1016/0370-2693(92)91241-z} {\bibfield  {journal} {\bibinfo  {journal}
  {Physics Letters B}\ }\textbf {\bibinfo {volume} {289}},\ \bibinfo {pages}
  {411--416} (\bibinfo {year} {1992})}\BibitemShut {NoStop}%
\bibitem [{\citenamefont {Andersen}\ and\ \citenamefont
  {Wilcox}(1997)}]{Andersen:1996qb}%
  \BibitemOpen
  \bibfield  {author} {\bibinfo {author} {\bibfnamefont {William}\ \bibnamefont
  {Andersen}}\ and\ \bibinfo {author} {\bibfnamefont {Walter}\ \bibnamefont
  {Wilcox}},\ }\bibfield  {title} {\enquote {\bibinfo {title} {{Lattice charge
  overlap. 1. Elastic limit of pi and rho mesons}},}\ }\href {\doibase
  10.1006/aphy.1996.5648} {\bibfield  {journal} {\bibinfo  {journal} {Annals
  Phys.}\ }\textbf {\bibinfo {volume} {255}},\ \bibinfo {pages} {34--59}
  (\bibinfo {year} {1997})},\ \Eprint {http://arxiv.org/abs/hep-lat/9502015}
  {arXiv:hep-lat/9502015} \BibitemShut {NoStop}%
\bibitem [{\citenamefont {Wilcox}(2002)}]{Wilcox_2002}%
  \BibitemOpen
  \bibfield  {author} {\bibinfo {author} {\bibfnamefont {Walter}\ \bibnamefont
  {Wilcox}},\ }\bibfield  {title} {\enquote {\bibinfo {title} {Continuum moment
  equations on the lattice},}\ }\href {\doibase 10.1103/physrevd.66.017502}
  {\bibfield  {journal} {\bibinfo  {journal} {Physical Review D}\ }\textbf
  {\bibinfo {volume} {66}} (\bibinfo {year} {2002}),\
  10.1103/physrevd.66.017502}\BibitemShut {NoStop}%
\bibitem [{\citenamefont {Draper}\ \emph {et~al.}(1989)\citenamefont {Draper},
  \citenamefont {Woloshyn}, \citenamefont {Wilcox},\ and\ \citenamefont
  {Liu}}]{Draper:1988bp}%
  \BibitemOpen
  \bibfield  {author} {\bibinfo {author} {\bibfnamefont {Terrence}\
  \bibnamefont {Draper}}, \bibinfo {author} {\bibfnamefont {R.~M.}\
  \bibnamefont {Woloshyn}}, \bibinfo {author} {\bibfnamefont {Walter}\
  \bibnamefont {Wilcox}}, \ and\ \bibinfo {author} {\bibfnamefont {Keh-Fei}\
  \bibnamefont {Liu}},\ }\bibfield  {title} {\enquote {\bibinfo {title} {{The
  Pion Form-factor in Lattice {QCD}}},}\ }\href {\doibase
  10.1016/0550-3213(89)90609-3} {\bibfield  {journal} {\bibinfo  {journal}
  {Nucl. Phys. B}\ }\textbf {\bibinfo {volume} {318}},\ \bibinfo {pages}
  {319--336} (\bibinfo {year} {1989})}\BibitemShut {NoStop}%
\bibitem [{\citenamefont {Wilcox}\ and\ \citenamefont
  {Anderson-Pugh}(1994)}]{Wilcox:1993uq}%
  \BibitemOpen
  \bibfield  {author} {\bibinfo {author} {\bibfnamefont {Walter}\ \bibnamefont
  {Wilcox}}\ and\ \bibinfo {author} {\bibfnamefont {B.}~\bibnamefont
  {Anderson-Pugh}},\ }\bibfield  {title} {\enquote {\bibinfo {title}
  {{Structure functions, form-factors, and lattice QCD}},}\ }\href {\doibase
  10.1016/0920-5632(94)90400-6} {\bibfield  {journal} {\bibinfo  {journal}
  {Nucl. Phys. B Proc. Suppl.}\ }\textbf {\bibinfo {volume} {34}},\ \bibinfo
  {pages} {393--395} (\bibinfo {year} {1994})},\ \Eprint
  {http://arxiv.org/abs/hep-lat/9312034} {arXiv:hep-lat/9312034} \BibitemShut
  {NoStop}%
\bibitem [{\citenamefont {Baral}\ \emph {et~al.}(2019)\citenamefont {Baral},
  \citenamefont {Whyte}, \citenamefont {Wilcox},\ and\ \citenamefont
  {Morgan}}]{BARAL201964}%
  \BibitemOpen
  \bibfield  {author} {\bibinfo {author} {\bibfnamefont {Suman}\ \bibnamefont
  {Baral}}, \bibinfo {author} {\bibfnamefont {Travis}\ \bibnamefont {Whyte}},
  \bibinfo {author} {\bibfnamefont {Walter}\ \bibnamefont {Wilcox}}, \ and\
  \bibinfo {author} {\bibfnamefont {Ronald~B.}\ \bibnamefont {Morgan}},\
  }\bibfield  {title} {\enquote {\bibinfo {title} {Disconnected loop
  subtraction methods in lattice qcd},}\ }\href {\doibase
  https://doi.org/10.1016/j.cpc.2019.03.011} {\bibfield  {journal} {\bibinfo
  {journal} {Computer Physics Communications}\ }\textbf {\bibinfo {volume}
  {241}},\ \bibinfo {pages} {64--79} (\bibinfo {year} {2019})}\BibitemShut
  {NoStop}%
\bibitem [{\citenamefont {Romero}\ \emph {et~al.}(2020)\citenamefont {Romero},
  \citenamefont {Stathopoulos},\ and\ \citenamefont {Orginos}}]{Romero_2020}%
  \BibitemOpen
  \bibfield  {author} {\bibinfo {author} {\bibfnamefont {Eloy}\ \bibnamefont
  {Romero}}, \bibinfo {author} {\bibfnamefont {Andreas}\ \bibnamefont
  {Stathopoulos}}, \ and\ \bibinfo {author} {\bibfnamefont {Kostas}\
  \bibnamefont {Orginos}},\ }\bibfield  {title} {\enquote {\bibinfo {title}
  {Multigrid deflation for lattice qcd},}\ }\href {\doibase
  10.1016/j.jcp.2020.109356} {\bibfield  {journal} {\bibinfo  {journal}
  {Journal of Computational Physics}\ }\textbf {\bibinfo {volume} {409}},\
  \bibinfo {pages} {109356} (\bibinfo {year} {2020})}\BibitemShut {NoStop}%
\bibitem [{\citenamefont {Liu}\ \emph {et~al.}(2018)\citenamefont {Liu},
  \citenamefont {Liang},\ and\ \citenamefont {Yang}}]{Liu:2017man}%
  \BibitemOpen
  \bibfield  {author} {\bibinfo {author} {\bibfnamefont {Keh-Fei}\ \bibnamefont
  {Liu}}, \bibinfo {author} {\bibfnamefont {Jian}\ \bibnamefont {Liang}}, \
  and\ \bibinfo {author} {\bibfnamefont {Yi-Bo}\ \bibnamefont {Yang}},\
  }\bibfield  {title} {\enquote {\bibinfo {title} {{Variance Reduction and
  Cluster Decomposition}},}\ }\href {\doibase 10.1103/PhysRevD.97.034507}
  {\bibfield  {journal} {\bibinfo  {journal} {Phys. Rev. D}\ }\textbf {\bibinfo
  {volume} {97}},\ \bibinfo {pages} {034507} (\bibinfo {year} {2018})},\
  \Eprint {http://arxiv.org/abs/1705.06358} {arXiv:1705.06358 [hep-lat]}
  \BibitemShut {NoStop}%
\bibitem [{\citenamefont {Morningstar}\ \emph {et~al.}(2011)\citenamefont
  {Morningstar}, \citenamefont {Bulava}, \citenamefont {Foley}, \citenamefont
  {Juge}, \citenamefont {Lenkner}, \citenamefont {Peardon},\ and\ \citenamefont
  {Wong}}]{Morningstar_2011}%
  \BibitemOpen
  \bibfield  {author} {\bibinfo {author} {\bibfnamefont {C.}~\bibnamefont
  {Morningstar}}, \bibinfo {author} {\bibfnamefont {J.}~\bibnamefont {Bulava}},
  \bibinfo {author} {\bibfnamefont {J.}~\bibnamefont {Foley}}, \bibinfo
  {author} {\bibfnamefont {K.~J.}\ \bibnamefont {Juge}}, \bibinfo {author}
  {\bibfnamefont {D.}~\bibnamefont {Lenkner}}, \bibinfo {author} {\bibfnamefont
  {M.}~\bibnamefont {Peardon}}, \ and\ \bibinfo {author} {\bibfnamefont
  {C.~H.}\ \bibnamefont {Wong}},\ }\bibfield  {title} {\enquote {\bibinfo
  {title} {Improved stochastic estimation of quark propagation with laplacian
  heaviside smearing in lattice qcd},}\ }\href {\doibase
  10.1103/PhysRevD.83.114505} {\bibfield  {journal} {\bibinfo  {journal} {Phys.
  Rev. D}\ }\textbf {\bibinfo {volume} {83}},\ \bibinfo {pages} {114505}
  (\bibinfo {year} {2011})}\BibitemShut {NoStop}%
\bibitem [{\citenamefont {Wilcox}(1993)}]{Wilcox:1992xe}%
  \BibitemOpen
  \bibfield  {author} {\bibinfo {author} {\bibfnamefont {Walter}\ \bibnamefont
  {Wilcox}},\ }\bibfield  {title} {\enquote {\bibinfo {title} {{Lattice charge
  overlap and the elastic limit}},}\ }\href {\doibase
  10.1016/0920-5632(93)90257-7} {\bibfield  {journal} {\bibinfo  {journal}
  {Nucl. Phys. B Proc. Suppl.}\ }\textbf {\bibinfo {volume} {30}},\ \bibinfo
  {pages} {491--494} (\bibinfo {year} {1993})},\ \Eprint
  {http://arxiv.org/abs/hep-lat/9211011} {arXiv:hep-lat/9211011} \BibitemShut
  {NoStop}%
\end{thebibliography}%
%%%%%%%%%%%%%%%%%%%%%%%%%%%%

\end{document}